\documentclass[letterpaper,twocolumn,10pt]{article}
\usepackage{usenix2019_v3}

\usepackage{tikz}
\usepackage{amsmath}

\usepackage[]{hyperref}
\usepackage{amsfonts}
\usepackage{url}
\usepackage{filecontents}
\usepackage{xspace}
\usepackage{subfig}
\usepackage{multirow}
\usepackage{makecell}
\usepackage[ruled, linesnumbered]{algorithm2e}

\usepackage[T1]{fontenc}

\usepackage{enumitem}

\begin{document}

\newcommand{\SystemName}{\textsc{SwiftDiffusion}\xspace}

\newcommand{\wei}[1]{{{\color{red}#1}}{}}
\newcommand{\suyi}[1]{{{\color{cyan}#1}}{}}
\newcommand{\lingyun}[1]{{{\color{brown}#1}}{}}

\newcommand*\circled[1]{\tikz[baseline=(char.base)]{\node[shape=circle,draw,inner sep=0.3pt] (char) {#1};}}

\newcommand{\PHB}[1]{\noindent\textbf{#1}\hspace{.5em}} 
\newcommand{\PHM}[1]{\vspace{.4em} \noindent\textbf{#1}\hspace{.5em}} 

\newcommand{\secref}[1]{\S\ref{#1}}
\newcommand{\figref}[1]{Fig.~\ref{#1}}
\newcommand{\tabref}[1]{Table~\ref{#1}}
\newcommand{\thmref}[1]{Theorem~\ref{#1}}
\newcommand{\prgref}[1]{Program~\ref{#1}}
\newcommand{\algref}[1]{Algorithm~\ref{#1}}
\newcommand{\eqnref}[1]{Eqn.~\ref{#1}}
\newcommand{\clmref}[1]{Claim~\ref{#1}}
\newcommand{\lemref}[1]{Lemma~\ref{#1}}
\newcommand{\ptyref}[1]{Property~\ref{#1}}

\newcommand{\eg}{{e.g.\@\xspace}}
\newcommand{\ie}{{i.e.\@\xspace}}
\newcommand{\etc}{
        \@ifnextchar{.}
        \textit{etc}
        \textit{etc.\@\xspace}
}

\newcommand{\term}{\textsf}
\newcommand{\code}{\texttt}
\newcommand{\ths}{\textsuperscript{th}}
\newcommand{\circledtext}[1]{\raisebox{.5pt}{\textcircled{\raisebox{-.9pt} {#1}}}}

\setlength{\abovecaptionskip}{3pt plus 1pt minus 1pt}
\setlength{\belowcaptionskip}{3pt plus 1pt minus 1pt}
\setlength{\abovedisplayskip}{3pt}
\setlength{\belowdisplayskip}{3pt}

\def\alibaba{{$^\diamond$}}

\date{}

\title{\Large \bf \SystemName: Efficient Diffusion Model Serving with Add-on Modules}


\author{
\rm Suyi Li$^*$, Lingyun Yang$^*$, Xiaoxiao Jiang, Hanfeng Lu, Dakai An, Zhipeng Di{\alibaba}, Weiyi Lu{\alibaba}, Jiawei Chen{\alibaba} \\ Kan Liu{\alibaba}, Yinghao Yu{\alibaba}, Tao Lan{\alibaba}, Guodong Yang{\alibaba}, Lin Qu{\alibaba}, Liping Zhang{\alibaba}, Wei Wang\\
HKUST, {\alibaba}Alibaba Group \\
}

\maketitle
\def\thefootnote{*}\footnotetext{Equal contribution}
\def\thefootnote{\arabic{footnote}}

\begin{abstract}

Text-to-image (T2I) generation using diffusion models has become a blockbuster
service in today's AI cloud. A production T2I service typically involves a
serving workflow where a base diffusion model is augmented with various
``add-on'' modules, notably ControlNet and LoRA, to enhance image generation
control. Compared to serving the base model alone, these add-on modules
introduce significant loading and computational overhead, resulting in
increased latency. In this paper, we present \SystemName, a system that
efficiently serves a T2I workflow through a holistic approach. \SystemName
decouples ControNet from the base model and deploys it as a \emph{separate,
independently scaled service on dedicated GPUs}, enabling ControlNet
caching, parallelization, and sharing. To mitigate the high loading overhead
of LoRA serving, \SystemName employs
a \emph{bounded asynchronous LoRA loading} (BAL) technique, allowing 
LoRA loading to overlap with the initial base model execution by \emph{up to $k$ steps} 
without compromising image quality. Furthermore, \SystemName optimizes base
model execution with a novel \emph{latent parallelism} technique. 
Collectively, these designs enable \SystemName to outperform the state-of-the-art 
T2I serving systems, achieving up to $7.8\times$ latency
reduction and $1.6\times$ throughput improvement in serving SDXL models on 
H800 GPUs, without sacrificing image quality.
\end{abstract}

\section{Introduction}
\label{sec:intro}

Text-to-image (T2I) generation using diffusion models is a transformative AI
technology that enables the creation of high-quality, contextually accurate
images from textual descriptions. This technology has gained immense
popularity, with a variety of prominent T2I services available in the cloud,
such as DALL·E~\cite{dall-e}, Midjourney~\cite{midjourney}, and Firefly~\cite
{firefly}. Notably, Firefly has reportedly generated over 2 billion
images~\cite{adobe_report}, highlighting the significant demand for T2I
services.

\begin{figure}[t]
  \centering
  \includegraphics[width=0.95\linewidth]{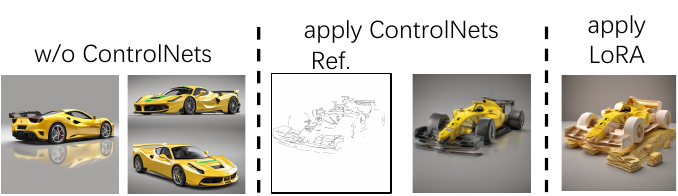}
  \caption{Effects of ControlNet and LoRA in image generation with SDXL under the same prompt: racing game car, yellow Ferrari. \textbf{Left}: without ControlNet, the generated images can have different compositions. \textbf{Center}: ControlNet uses a reference image to control the composition. \textbf{Right}: using LoRA to generate image in a papercut style.}
  \label{fig:addon_effect}
  \vspace{-.15in}
\end{figure}

A production T2I service typically consists of multiple components. At its
core is a base stable diffusion model~\cite{podell2024sdxl, sd3,li2024hunyuan}. This model is trained to produce a coherent 
image through a \emph{reverse diffusion process}~\cite{clip}: it starts with an image composed of
random noises and progressively denoises this random input in iterations,
until the output image aligns with the provided text description. In addition
to the base diffusion model, production T2I services provide various
``add-on'' modules, including ControlNet~\cite{zhang2023controlnet} and LoRA
(Low-Rank Adaptation~\cite{hu2022lora}), that allow users to control the details of
the output images, such as shapes, outlines, poses, and styles. \figref
{fig:addon_effect} illustrates the effects of using these two modules.
ControlNet allows users to input a reference image to guide the spatial
composition of the generated image, while LoRA produces an image with
customized stylistic effects. In our production platform, over $98\%$ of
requests demand at least one ControlNet, and over $95\%$ utilize at least one LoRA
(\secref{sec:characterization_study}).

\begin{figure}[t]
  \centering
  \includegraphics[width=0.8\linewidth]{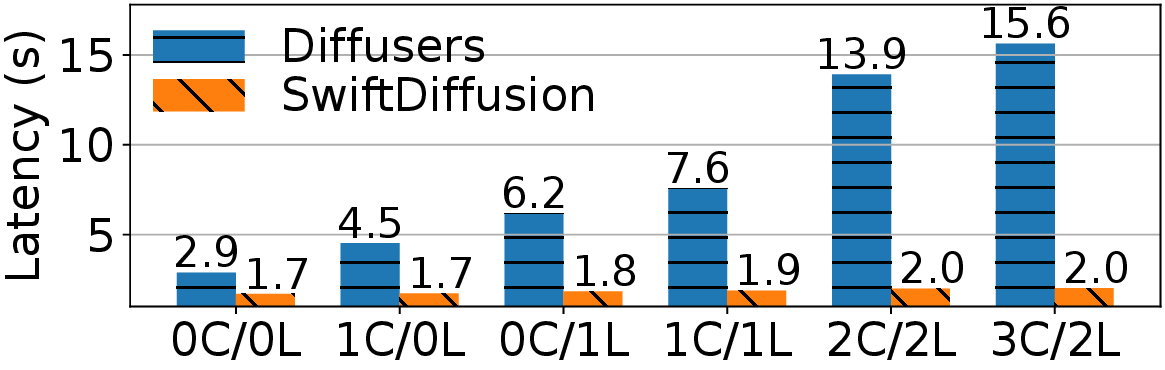}  
    \caption{ \textbf{C}ontrolNets and \textbf{L}oRAs introduce additional latency overhead. In each workflow, a common
    base SDXL~\cite{podell2024sdxl} model is augmented with
    $m$ ControlNets and $n$ LoRAs ($m$C/$n$L), served by Diffusers~\cite{diffusers} on an H800 GPU.
    }
  \label{fig:lat_multi_controlnets_lora_gpu_saturation}
  \vspace{-.2in}
\end{figure}

The base diffusion model, together with the requested add-on modules,
compose a \emph{T2I serving workflow}. However, the use of these add-on modules poses
new challenges. To illustrate this, we configure a T2I workflow where a
base SDXL model~\cite{podell2024sdxl} is augmented with a varying number
of ControlNets and LoRAs. \figref
{fig:lat_multi_controlnets_lora_gpu_saturation} depicts the serving
latency of these workflows (blue bar). Compared with serving the base
model alone (0C/0L), serving it together with add-on modules results in a
significant delay, which is increasing as more ControlNets and LoRAs are
in use. This delay mainly comes from two sources. \textbf{First}, as the desired
ControlNets and LoRAs vary across requests, they must be fetched from storage
and loaded into GPU memory before serving, introducing significant loading
overhead. In our production platform, each request undergoes one ControlNet
loading and one LoRA loading on average, which account for $37\%$ of the
end-to-end serving latency (\secref{sec:characterization_study}). Note that pre-caching all add-on
modules in GPU memory is \emph{infeasible}: our production trace reports nearly 150 distinct
ControlNets and 14,500 LoRAs requested by users; each ControlNet
is around 3 GiB and each LoRA is hundreds of MiB\footnote{The ControlNets and
LoRAs are for the SDXL model.}. \textbf{Second}, while LoRA is lightweight, 
ControlNet is compute-intensive. 
\figref{fig:lat_multi_controlnets_lora_gpu_saturation} shows that using one
ControlNet in T2I generation adds 1.6 seconds to the serving latency, which
is $1.5\times$ longer than serving the base model alone (2.9 seconds). As
more ControlNets are utilized, their computational overhead accumulates, leading to a significant
latency increase (\figref{fig:lat_multi_controlnets_lora_gpu_saturation}).

Despite these challenges, efficiently serving a T2I workflow with add-on
modules has been largely unexplored; prior work primarily focuses on
improving the serving latency and image quality of a single base diffusion
model~\cite{diffusers, nirvana, wangDiffusionDBLargescalePrompt2023,
li2024DistriFusion}. In this paper, we propose \SystemName,
a
system that efficiently serves a base diffusion model and the associated
ControlNets and LoRAs for enhanced image generation control. \SystemName employs a
holistic approach with three novel designs driven by a characterization study
in a production platform (\S\ref{sec:characterization_study}):

\PHM{ControlNet-as-a-Service.}
Efficient ControlNet serving requires addressing the GPU loading and
computational overhead. Our characterization study reveals that ControlNets exhibit
the \emph{skewed popularity}; that is, a small number of ControlNets
(9--11\%) are invoked frequently by a large number of user requests
(95--98\%). Caching these popular ControlNets in GPU memory effectively
reduces the loading tax, with only modest memory footprint. To accelerate
computation, \SystemName \emph{concurrently executes} ControlNet(s) with the
base diffusion model on \emph{multiple GPUs}, achieving \emph
{close-to-ideal speedup} compared to the current sequential execution
schemes~\cite{diffusers}.

\SystemName implements ControlNet caching and parallelization with a new
 design called \emph{ControlNet-as-a-Service}. It decouples ControlNets from
 the base model and deploys them as a \emph{separate, independently
 scaled service on dedicated GPUs}, where popular ControlNets are
 cached in GPU memory to eliminate the loading overhead. This service can be
 dynamically invoked to execute desired ControlNets in
 parallel with the base model. This design additionally
 enables \emph{ControlNet sharing}, in that a single ControlNet can be
 multiplexed by multiple base models.

\PHM{Bounded asynchronous LoRA loading.}
Unlike ControlNet, LoRA is compute-light and LoRA serving is mainly
bottlenecked by the loading overhead. Given their large populations, 
LoRA adapters are usually maintained in disk or remote memory storage and must be
brought into GPU memory on-demand. We notice that LoRA caching offers limited
benefits in this scenario as LoRA adapters exhibit a \emph
{heavy-tailed distribution in popularity} (\S\ref
{sec:characterization_lora}). 

To address this challenge, we analyze the T2I generation process with and
without LoRAs and find that the two computations diverge largely in the \emph
{later stage} of the denoising process. This suggests that one can exclude LoRA
computation in initial steps and include them only in latter iterations,
while still producing images of the same quality. Based on this insight, we
propose \emph{bounded asynchronous LoRA loading} (BAL). That is, while the
requested LoRAs are being loaded into GPUs, \SystemName asynchronously
executes the base model (and the desired ControlNets, if any) to early
start the image generation process without LoRA by \emph{up to} $k$ steps, beyond which
LoRA computation must be included to continue the
remaining generation steps. By tuning the asynchronous bound $k$, \SystemName effectively overlaps
LoRA loading with base model execution, without compromising image quality
(\S\ref{sec:eval_performance}).

\PHM{Latent parallelism for CFG computation.}
In addition to add-on modules, \SystemName optimizes base model execution with
a new parallelism technique. As mentioned earlier, T2I generation is
essentially a denoising process, where the diffusion model progressively
refines a \emph{latent tensor}, initially filled with random noise, through
multiple denoising steps~\cite{podell2024sdxl, li2024hunyuan, sd3}. Each
denoising step employs the classifier-free guidance (CFG) to effectively
improve image quality and alignment~\cite{ho2021classifierfree}. In CFG, the
input latent tensor is duplicated and the two replicas undergo two different
denoising processes, one guided by the text prompt (conditioned denoising)
and the other not (unconditioned denoising). The two denoised tensors are
then aggregated by computing a weighted sum. Since conditioned and
unconditioned denoising have no dependency, \SystemName parallelizes the two
computations on two GPUs. This \emph{latent parallelism} technique can also 
be applied to accelerate the CFG
computation in ControlNet serving. This technique, together with other
engineering optimizations, collectively accelerate base model execution by
$1.7\times$ (\secref{sec:microbenchmark_unet}).


We have implemented \SystemName on top of HuggingFace Diffusers~\cite
{diffusers} and evaluated its performance using text prompts from
PartiPrompts~\cite{partiprompt}. Our evaluation encompasses SDXL~\cite
{podell2024sdxl}, a UNet-based diffusion model~\cite{UNet} widely deployed in
production~\cite{nirvana,kolors}, and two recently proposed diffusion
transformer (DiT) models~\cite{peebles2023scalable, sd3, li2024hunyuan}.
Evaluation results demonstrate that \SystemName outperforms Nirvana~\cite
{nirvana}, the state-of-the-art T2I serving system, reducing the average
serving latency by up to $7\times$, improving the throughput by up to
$1.5\times$, and generating images of better quality (\secref
{sec:eval_performance}). To comprehensively assess the quality of the
generated images, we engaged 75 human users. Their assessment report
confirms that \SystemName produces images of the same quality as
Diffusers~\cite{diffusers}. Our contributions are summarized as follows:

\begin{itemize}[topsep=5pt, leftmargin=*, noitemsep, nolistsep, parsep=5pt, partopsep=0pt]

\item We present the first characterization study in a production T2I
 platform and identify new challenges of serving base diffusion models
 with add-on modules. 

\item We propose a holistic approach with three novel designs to
 systematically optimize the T2I serving workflow, including
 ControlNet-as-a-Service, bounded asynchronous LoRA loading, and latent
 parallelism for CFG computation.  

 \item We develop \SystemName, an optimized serving system for T2I 
 applications that achieves significant speedup without
 compromising image quality (\figref{fig:eval_real_examples}).

\end{itemize}

\section{Background}
\label{sec:background}

In this section, we give a primer to diffusion-based T2I generation
and the use of two add-on modules, namely ControlNet and LoRA, for enhanced
generation control.

\begin{figure}[t]
  \centering
  \includegraphics[width=0.99\linewidth]{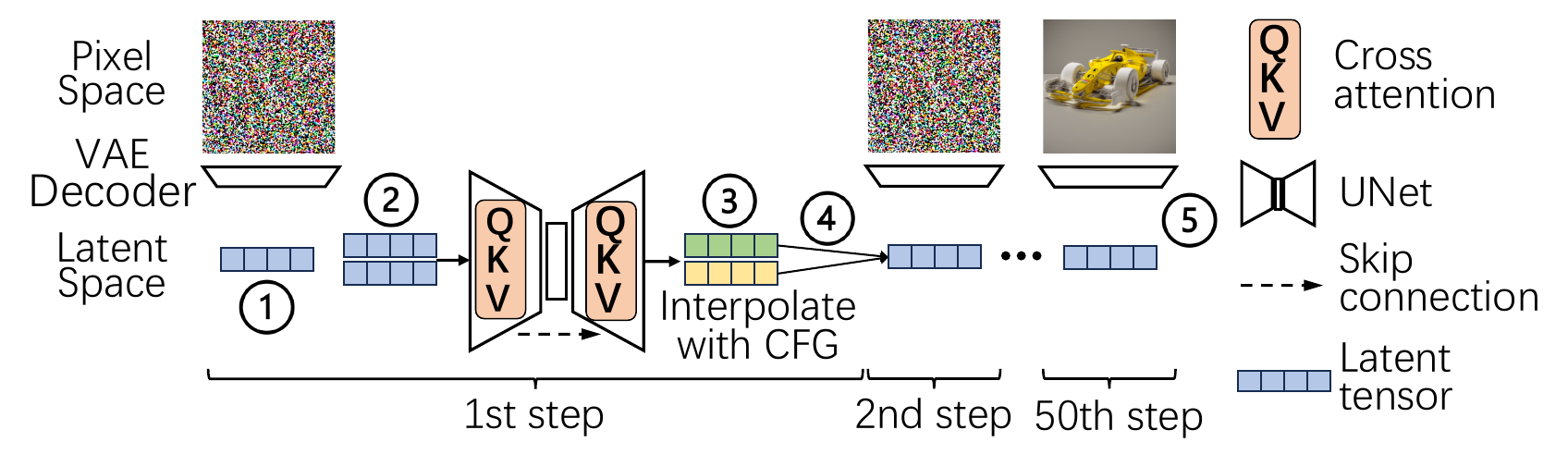}
  \caption{A workflow of text-to-image with a stable diffusion model. Time embedding is ignored for simplicity.}
  \label{fig:sdxl_inference}
  \vspace{-.2in}
\end{figure}

\PHM{Diffusion model.}
A typical stable diffusion model~\cite{rombach2022high, podell2024sdxl}
consists of three main components: a text encoder~\cite{clip}, a
convolutional UNet model~\cite{UNet}, and a decoder-only variational
autoencoder (VAE). The model generates images through a denoising process illustrated in 
\figref{fig:sdxl_inference}.
Given a text prompt, the text encoder encodes the prompt into token
embeddings. The image generation process then begins by initializing a \emph
{latent tensor} filled with \emph{random noise} (\circled{1}), which is progressively
refined by the UNet over multiple denoising steps guided by the token embeddings. To steer the image generation
towards the desired outcome, the UNet employs classifier-free guidance~\cite
{ho2021classifierfree} (CFG). Specifically, at each denoising step, the UNet
duplicates the latent tensor into two replicas (\circled{2}); one replica is denoised \emph
{conditionally}, taking into account the token embeddings, whilst the other is
denoised \emph{unconditionally}. Intuitively, the unconditioned latent representation
captures the general image distribution, whereas the conditioned representation 
incorporates specific context given by the text prompt (\circled{3}).
The two latent tensors are then combined by computing a weighted sum,
yielding an \textit{interpolated} latent representation for further refinement in the next
step (\circled{4}). Upon completion of the denoising process, the final interpolated latent
tensor is sent to the VAE decoder to render the output
image (\circled{5}). \figref{fig:sdxl_inference} illustrates this image generation process,
where the initial noisy latent tensor undergoes 50 denoising
steps to produce the final image.

\PHM{Serving objectives.}
Commercial diffusion-based T2I services typically have two serving
objectives. First, they should provide low latency to allow real-time user
interaction to better support multi-round prompt editing and image
fine-tuning~\cite{nirvana, dall-e}. Second, it is crucial to generate
high-quality images based on users' specifications on image composition,
color palette, poses, and artistic styles. This necessitates augmenting the
base diffusion model with add-on modules, including ControlNet and LoRA.

\PHM{ControlNet.}
As image generation is a stochastic process, users often find it
challenging to control this process, because text prompts alone are insufficient to
precisely specify complex layouts, compositions, and shapes. ControlNet~\cite{zhang2023controlnet}
addresses this issue by augmenting base diffusion models with 
additional input conditions, such as edge maps and
depth maps, which specify the desired spatial composition of the generated
images. As illustrated in \figref{fig:addon_effect}-Center, ControlNet
enables users to provide a reference edge map, allowing the base model to
generate a Ferrari that adheres to the specified spatial composition. In
practice, users can combine \textit{multiple} ControlNets for a single
image generation (\secref
{sec:characterization_controlnet}).

\figref{fig:controlnet_parallelization}-Left illustrates a simplified workflow of
 applying a ControlNet in the image generation process.
 ControlNet has a similar architecture to the UNet encoder blocks and
 middle block, with additional zero convolution operators. It is applied to
 each encoder level of the UNet backbone. In each denoising step (\figref
 {fig:sdxl_inference}), ControlNet takes as input the text prompt, the
 encoded reference image, and the latent tensor. The outputs,
 which contain the processed features of the reference image, are then incorporated
 into the skip-connections and middle block of the UNet backbone,
 guiding the image generation process to conform to the reference image. When 
 applying multiple ControlNets to a single base model, the
 outputs of these ControlNets are simply summed up and applied to the corresponding
 backbone UNet blocks~\cite{zhang2023controlnet}.


\PHM{Low-Rank Adaptation (LoRA).}
In addition to ControlNet, users utilize LoRA to generate images in a
customized style, as illustrated in \figref{fig:addon_effect}-Right. LoRA is
a parameter-efficient approach to enhancing the base model performance for
domain-specific tasks~\cite{hu2022lora}. It modifies a small subset of base
model parameters by patching new parameters to take effect. Specifically,
given a pre-trained weight matrix $\mathbf{W} \in \mathrm{R}^
{H_1 \times H_2}$ in a base model, LoRA introduces two low-rank matrices
$\mathbf{A} \in \mathrm{R}^{H_1 \times r } $ and  $\mathbf{B} \in \mathrm{R}^
{r \times H_2 }$, where $r$ is the LoRA rank. By modifying the weight matrix
to $\mathbf{W^{\prime}} = \mathbf{W} + \mathbf{A}\mathbf{B}$, LoRA
effectively stylizes the final generated image, infusing it with the desired
visual characteristics. 

\PHM{Other add-on modules.}
Our work primarily focuses on ControlNet~\cite{zhang2023controlnet} and LoRA~\cite{hu2022lora}, two widely adopted add-on modules in our production platform (\secref{sec:characterization_study}).
Meanwhile, there are emerging add-on modules~\cite{ju2024BrushNet, ye2023IP-Adapter, wang2024InstantID, Fooocus, IC-Light} introduced by the research community to control image generation, most of which are developed based on UNet architectures~\cite{sd3, podell2024sdxl, rombach2022high}.
These modules can be broadly categorized into two groups.
The first group comprises models that operate in tandem with the base model during inference, akin to ControlNet.
Representative examples include IP-Adapter~\cite{ye2023IP-Adapter}, InstantID~\cite{wang2024InstantID}, and BrushNet~\cite{ju2024BrushNet}.
The second group consists of models that augment the base model by incorporating parameter-efficient patches, similar to LoRA.
Examples include Fooocus Inpaint~\cite{Fooocus} and IC-Light~\cite{IC-Light}.
Notably, the techniques proposed in \SystemName are generalizable and can be adapted to support these alternative add-on modules.

\section{Characterization Study}
\label{sec:characterization_study}

In this section, we present a characterization study on a 20-day workload
trace collected in May and June 2024 on a production platform. The trace
contains more than 500k inference requests to two core T2I services for online
retailing applications.\footnote{We are working on releasing the trace for
public access.} Our characterization not only reflects the
deployment scenarios of diffusion models in production, but also reveals the
inefficiency of current T2I serving systems. 


\begin{table}[t]
    \footnotesize
    \centering
    \begin{tabular}{c c c c}
        \hline
        \textbf{Add-on Module}             & \textbf{Number} & \textbf{Service A}    & \textbf{Service B} \\
        \hline
        \multirow{4}*{ControlNet} & 0      & 0            & 1.9\% \\
                                  & 1      & 30.5\%       & 25.1\% \\
                                  & 2      & 69.5\%       & 69.9\%\\
                                  & 3      & 0            & 3.1\% \\
        \hline
        \multirow{3}*{LoRA}       & 0      & 0.2\%        & 7.2\% \\
                                  & 1      & 8.8\%        & 73.6\% \\
                                  & 2      & 91\%         & 19.2\% \\
        \hline
    \end{tabular}
    \caption{The distribution of the number of ControlNets and LoRAs used 
    by each request in two production services.}
    \vspace{-.1in}
    \label{tab:addon_usage}
\end{table}

\subsection{ControlNet Characterization}
\label{sec:characterization_controlnet}

\PHB{Prevalence.}
\tabref{tab:addon_usage} shows the distribution of the number of ControlNets
 utilized by each request in two services. ControlNet is used by almost all
 requests for image generation control; approximately 70\% of these requests
 utilize two or more ControlNets simultaneously.

\begin{figure}[tb]
  \centering
  \includegraphics[width=0.495\linewidth]{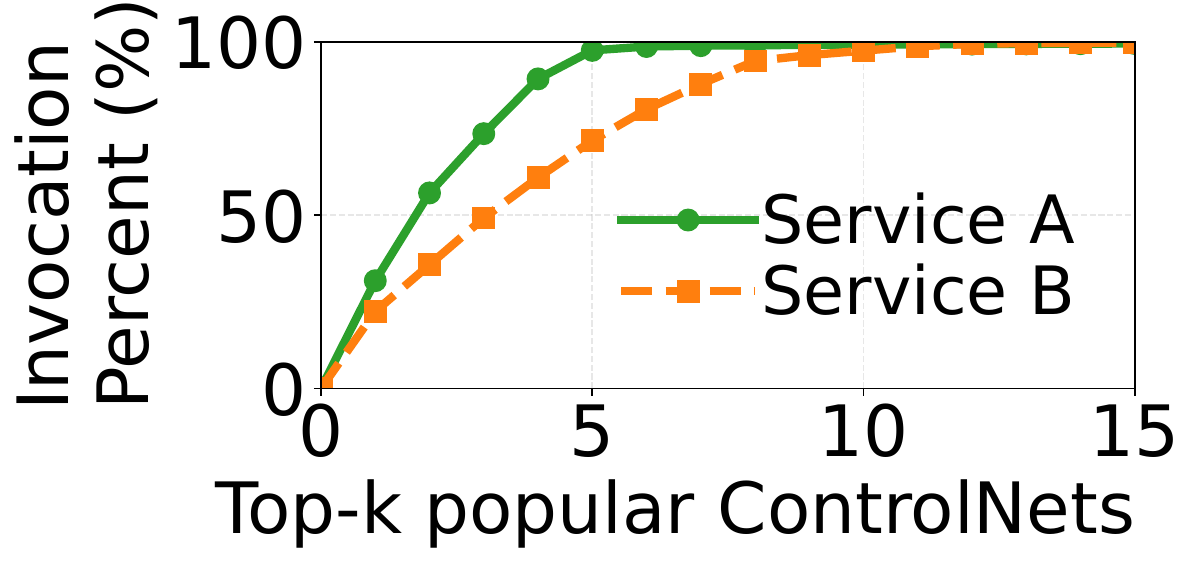}
  \hfill
  \centering
  \includegraphics[width=0.495\linewidth]{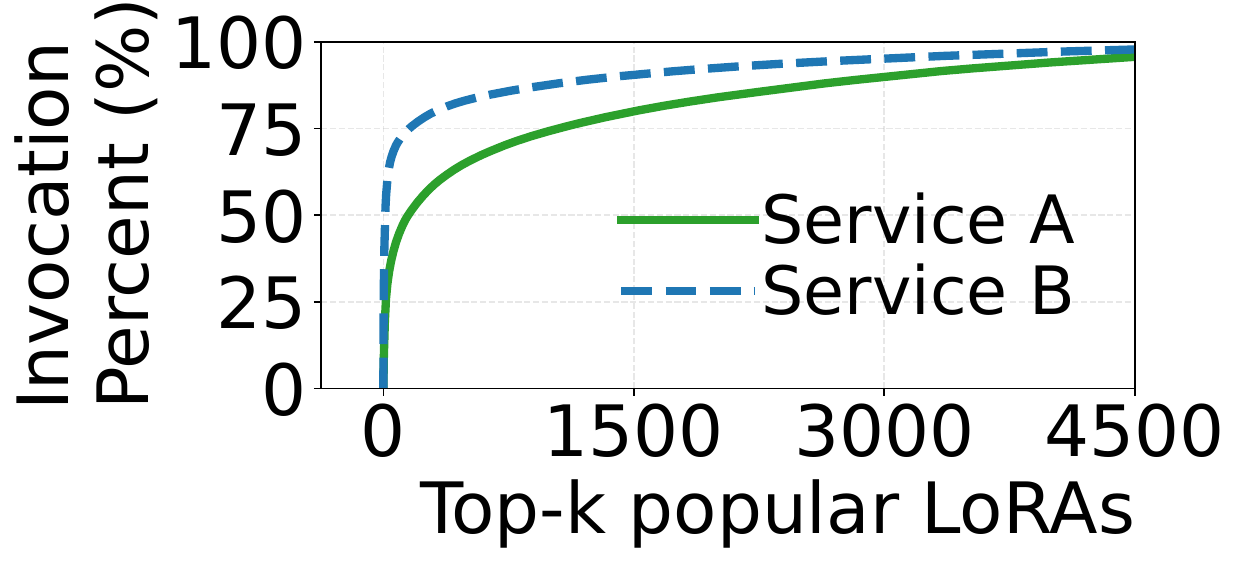}
  \caption{\textbf{Left}: ControlNet has a small population and exhibits a skewed popularity; the long tail of the graph is truncated for a better presentation. \textbf{Right}: LoRA has a large quantity and exhibits a long-tailed distribution in popularity.}
  \label{fig:addon_id_cdf}
  \vspace{-.1in}
\end{figure}

\PHM{Skewed popularity.}
Compared to a large quantity of requests, only 141 ControlNets
are used in two services, where Service A offers 47 distinct
ControlNets and Service B provides 94. These ControlNets exhibit a severe
imbalance in access frequency, with a handful of ControlNets being extremely
popular (\figref{fig:addon_id_cdf}-Left). In Service A, the top-5
popular ControlNets (11\% in population) account for 98\% of total
invocations; in Service B, the top-8 popular ControlNets (9\% in population) 
contribute to 95\% of total invocations.

\begin{figure}[tbp]
  \centering
  \includegraphics[width=0.495\linewidth]{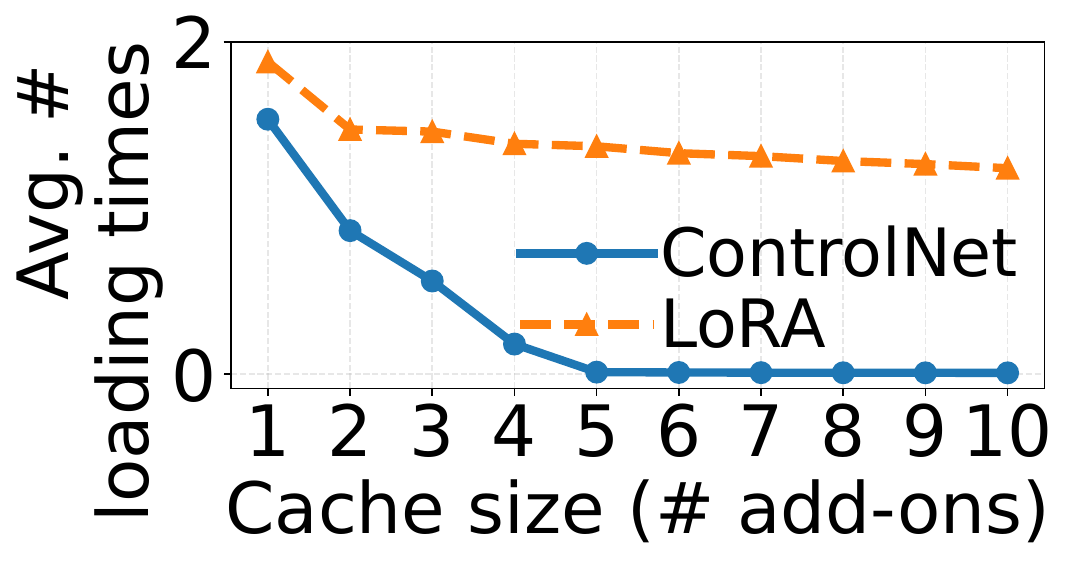}
  \hfill
  \centering
  \includegraphics[width=0.495\linewidth]{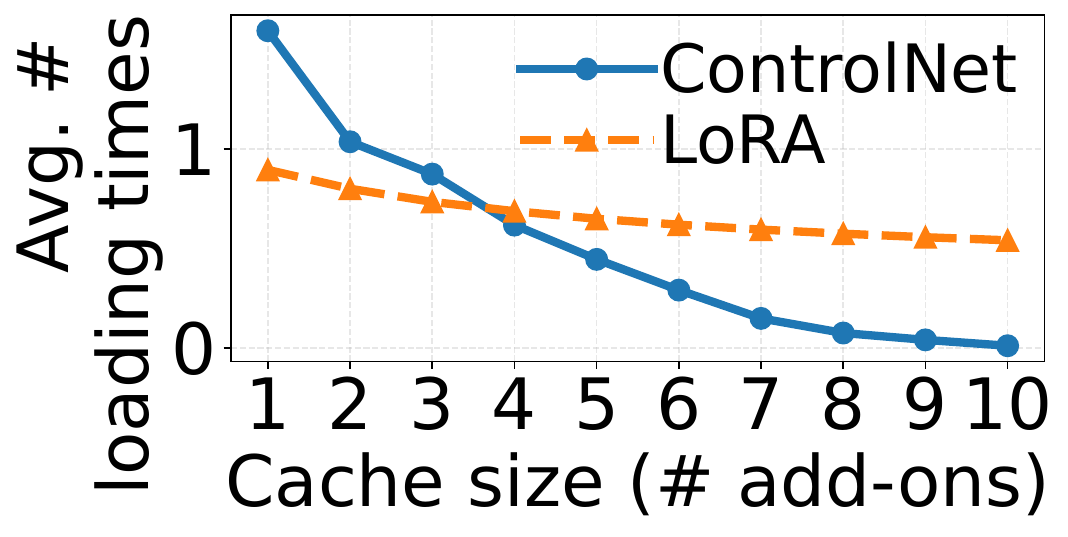}
  \caption{ControlNet loading overhead can be alleviated using a larger LRU cache, while LoRA performance gains are less pronounced. \textbf{Left}: Service A; \textbf{Right}: Service B.}
  \label{fig:addon_switch}
  \vspace{-.1in}
\end{figure}

\PHM{The need for ControlNet caching.}
ControlNets are large in size (3 GiB each) and usually maintained in remote
storage, introducing significant loading overhead. Given that ControlNets
have a limited quantity and skewed popularity, caching top popular
ControlNets in GPU memory can effectively reduce the loading overhead. To
illustrate this, we configure an LRU cache of varying size for ControlNet
caching. We replay the trace and measure the average number of times that the
desired ControlNets are not resident on GPU and must be fetched from storage
(i.e., cache miss) when serving two consecutive requests that desire
different ControlNets. As illustrated in \figref{fig:addon_switch}
(blue curves), caching only a handful of top popular ControlNets is
sufficient to eliminate the loading overhead (top-5 for Service A and top-8 for Service B).

\PHM{Computational overhead.}
ControlNet is compute-intensive as it shares a similar architecture to the
UNet encoder and middle block (\S\ref{sec:background}). As illustrated
in \figref{fig:lat_multi_controlnets_lora_gpu_saturation} (blue bars),
augmenting the base diffusion model with one ControlNet increases the serving
latency by $1.55\times$, up from 2.9 seconds to 4.5 seconds. As more
ControlNets are utilized by a request, their computational overhead
accumulates. This is primarily due to the inefficient \emph
{sequential} ControlNet execution design in current T2I serving
systems~\cite{diffusers}: at each denoising step, the system sequentially
computes the outputs of all the requested ControlNets before executing
the base model (\figref{fig:controlnet_parallelization}-Left).

\subsection{LoRA Characterization}
\label{sec:characterization_lora}

\PHB{Prevalence.} 
Similar to ControlNet, the vast majority of T2I requests utilize one or two
LoRAs to stylize the generated image, as shown in \tabref
{tab:addon_usage}. Specifically, over 90\% of requests in Service~A
desire two LoRAs, while nearly 74\% of requests in
Service B demand one LoRA.

\begin{figure}[t]
  \centering
  \includegraphics[width=0.495\linewidth]{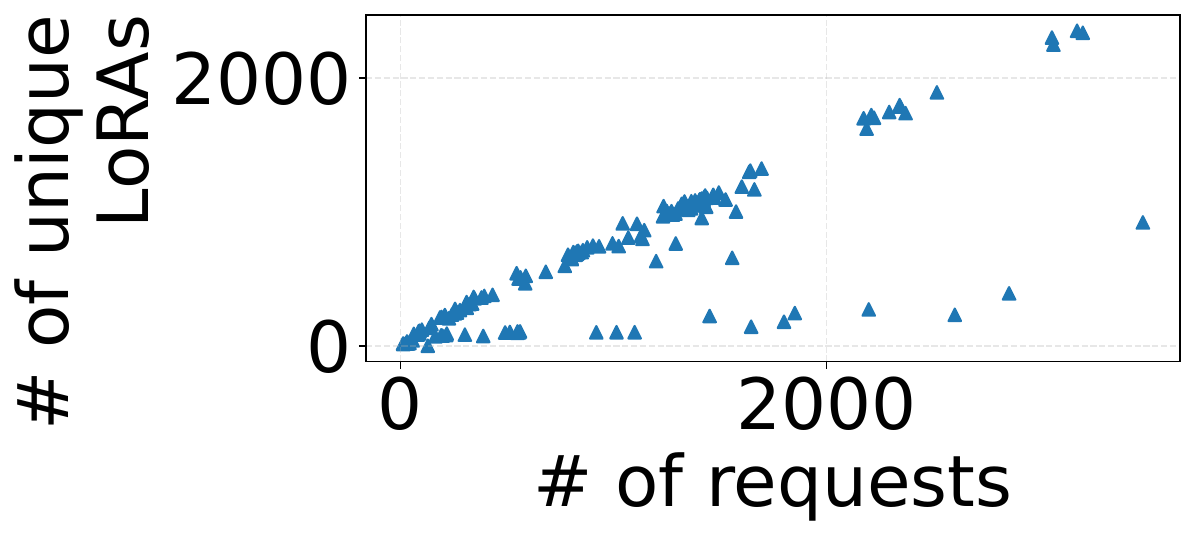}
  \hfill
  \centering
  \includegraphics[width=0.495\linewidth]{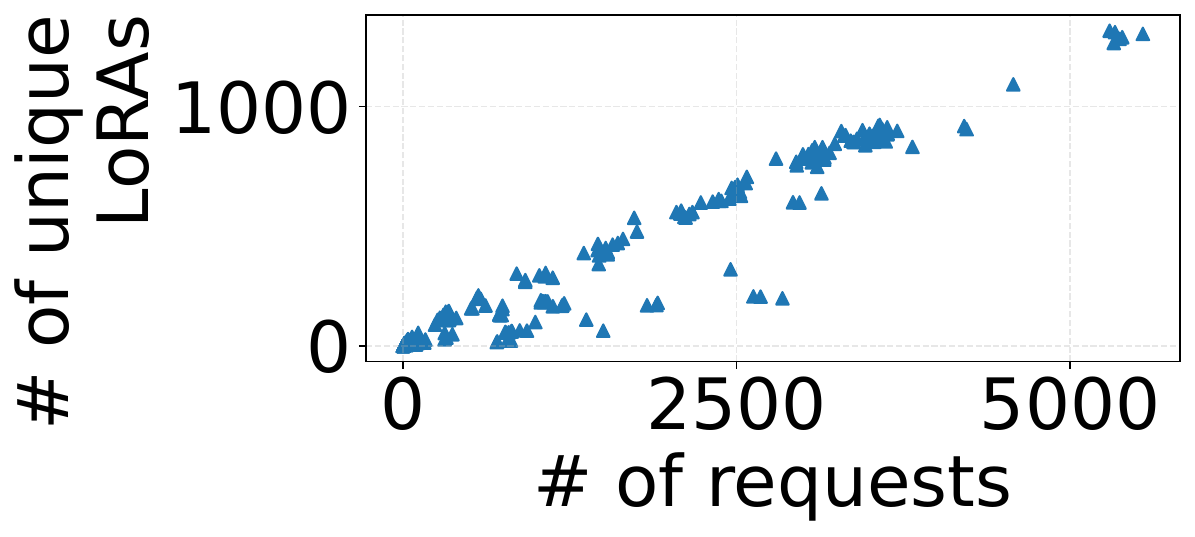}
  \caption{Scatter plots illustrating the number of requests received on each worker node (X-axis) against the number of unique LoRAs required by those requests (Y-axis). \textbf{Left}: Service A; \textbf{Right}: Service B.}
  \label{fig:unique_addon}
  \vspace{-.2in}
\end{figure}

\PHM{Long-tailed popularity.}
Compared to ControlNets, LoRAs have a significantly larger quantity but much
smaller sizes. Our trace reports 6,908 distinct LoRAs for Service A and 7,463
LoRAs for Service B. Each LoRA is a few hundreds of MiB. Unlike ControlNets,
the popularity of LoRAs follows a \emph{long-tailed distribution} in both
services; that is, a significant portion of invocations are contributed by a
large number of \emph{less popular} adapters, as illustrated in \figref
{fig:addon_id_cdf}-Right,

\PHM{Ineffective LoRA caching.}
Given the long-tailed popularity distribution of LoRAs, caching the top
popular adapters in GPU memory offers \emph{limited benefits}. To demonstrate
this, we configure an LRU cache of varying size for LoRA caching. We replay
the trace and measure the average number of times that the desired LoRAs are
not available on GPU and must be brought from storage (i.e., cache miss) when
serving two consecutive requests that demand different sets of LoRAs. As
illustrated in \figref{fig:addon_switch} (orange curves), configuring a
larger cache for LoRAs only results in a slight reduction of the loading
overhead caused by cache miss. As another evidence, we analyze the production
trace and depict in \figref{fig:unique_addon} a scatter plot illustrating the
number of requests served on each node and the number of \emph{unique} LoRAs
required by these requests. We observe the linear correlation
between the two numbers, invalidating the benefits of LoRA caching. As a
result, production systems do not cache LoRAs but on-demand load them from
local disk or a remote memory store, backed by a stable storage that manages LoRA
weight files.

\PHM{LoRA loading and patching overhead.}
Compared to ControlNets, LoRAs are compute-light and usually bottlenecked by
the loading and patching overhead. Our measurements show that fetching two
LoRAs with total size of approximately 800 MiB from a remote distributed
cache takes more than one second. This overhead alone delays the base
model serving by 34\% (up from 2.9 seconds to 3.9 seconds). In addition, simply
patching LoRA weights to the base model, as implemented in existing systems~\cite
{diffusers,peft}, can be extremely inefficient, which we elaborate
in \S\ref{sec:design_lora}.

\subsection{Characterizing Base Model Serving}
\label{sec:characterization_base_model}

Currently, UNet-based diffusion models, such as SDXL~\cite
{podell2024sdxl}, are predominately deployed to handle the majority of
requests in production T2I services. These models are supported by a plethora of
well-trained ControlNets~\cite{zhang2023controlnet} and LoRAs~\cite
{hu2022lora}. In the meantime, there is an emerging trend of deploying
transformer-based diffusion models~\cite{sd3,li2024hunyuan} for improved
image quality. As these models are recently developed, corresponding add-on
modules remain lagging behind, with limited choices and availability.
In this paper, we primarily focus on UNet-based models; our
observations and optimization designs also apply to the transformer backbone 
(\S\ref{sec:design_dit}).

\begin{figure}[t]
  \centering
\includegraphics[width=0.495\linewidth]{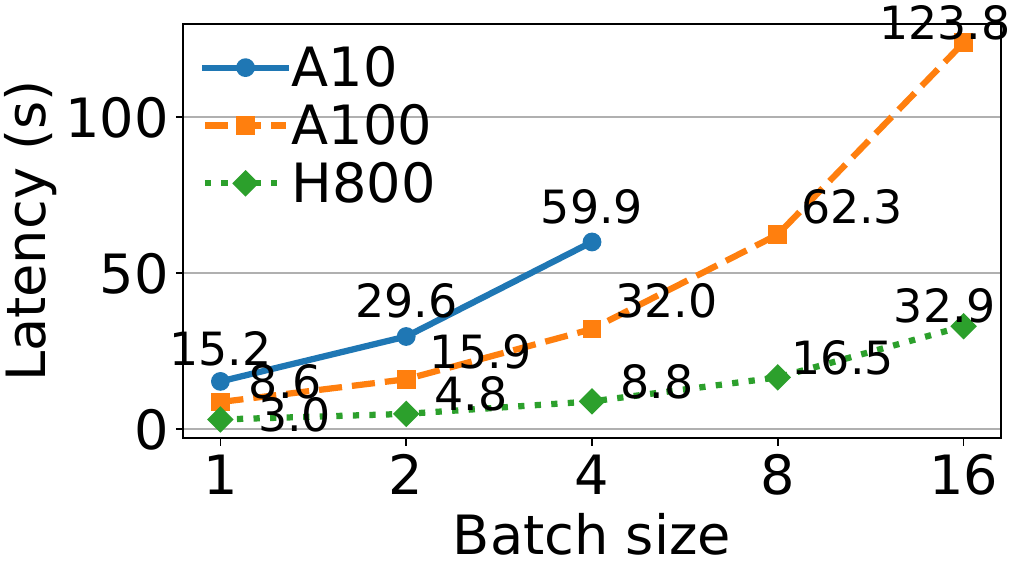}
  \includegraphics[width=0.495\linewidth]{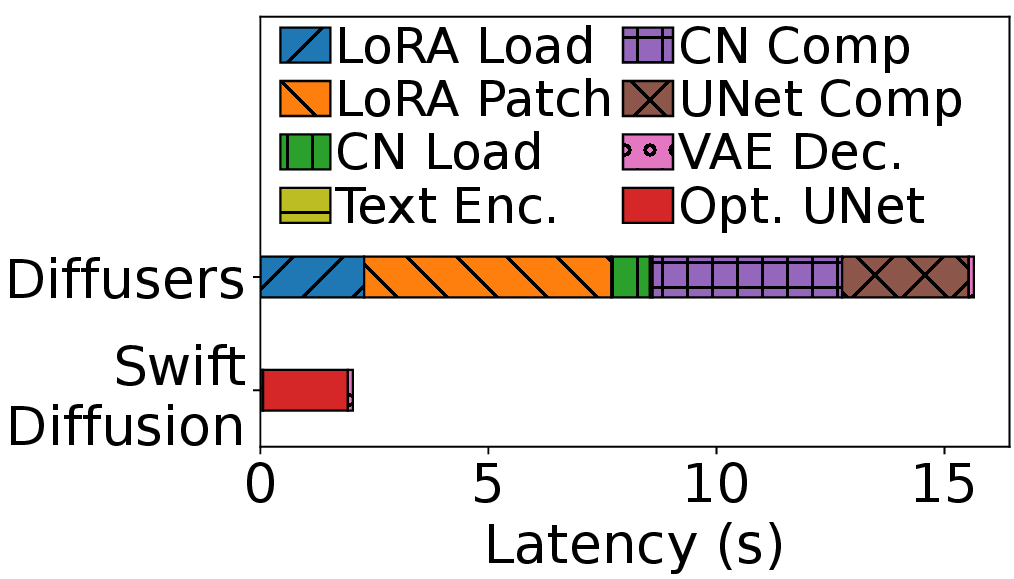}
  
  \caption{\textbf{Left}: SDXL inference saturates GPU. \textbf{Right}: Latency breakdown for a request with 3\textbf{C}/2\textbf{L}. \textbf{CN}: ControlNet.}
  \label{fig:saturation_latency_breakdown}
  \vspace{-.2in}
\end{figure}

\PHM{Limited benefits from batching.}
Diffusion model serving is computationally intensive, as evidenced by our
experiments with varying batch sizes for an SDXL model~\cite
{podell2024sdxl} on NVIDIA A10, A100, and H800 GPUs (\figref
{fig:saturation_latency_breakdown}-Left). Across all three GPUs, doubling the
batch size results in an approximately 2$\times$ in serving latency,
indicating minimal benefits from batching. In fact, generating a single image
already saturates the computational resources of a high-end GPU.
Consequently, production T2I services typically configure a constant batch
size of 1 to minimize serving latency.

\PHM{CFG dominates computation.}
To understand the computation of the base diffusion model, we break down
its execution and find that over 90\% of the execution time is spent on
CFG computation (\S\ref{sec:background}). Current CFG
implementation employs \emph{latent batching}. That is, in
each denoising step, the latent tensor is duplicated and the two replicas are
fed into the base model to perform conditional and unconditional denoising
operations in \emph{one batch}. However, as the two denoising operations are
by nature compute-heavy, batching them together yields minimal benefits,
leading to the similar performance with sequential execution even on
a high-end GPU. In fact, latent batching results in up to
1.7$\times$ slowdown in base model serving compared to an optimized design (\S\ref{sec:microbenchmark_unet}).

\subsection{Inefficiency of Current Serving Pipeline}
\label{sec:char_analytical_modeling}

To sum up, current T2I systems serve the base model and the
requested add-on modules in a \emph{sequential execution
pipeline}. Specifically, assume a request utilizing $m$ ControlNets and $n$
LoRA adapters. Upon request arrival, the system loads all the desired
ControlNets and LoRAs into GPU memory, followed by patching the $n$ LoRAs to
the base model. The system then encodes the text prompt and proceeds to the
denoising process in $N$ steps. At each step, it sequentially executes the $m$
ControlNets and the LoRA-patched base model to generate a latent representation. The final
latent representation is then sent to the VAE decoder to generate the output
image. The end-to-end request serving latency is hence given by Eq.~\eqref{eq:latency},
where the notations are defined in \tabref{tab:math_symbols}:
\begin{align}
  \vspace{-.2in}
  \label{eq:latency}
  \footnotesize
  \begin{split}
     T = & \underbrace{ T_{\texttt{Load}}^{ m\mathrm{C} } + T_{\texttt{Load}}^{ n\mathrm{L} } + T_{\texttt{Patch}}^{ n\mathrm{L} }}_{\substack{\text{time to load and patch} \\ \text{add-on modules}}} + \underbrace{T_{\texttt{Enc}} + \sum_{i=1}^{N}( T_{\texttt{Comp}}^{ m\mathrm{C} } + T_{\texttt{Comp}}^{B}) + T_{\texttt{VAE}}.}_{\text{computation time}}
  \end{split}
  \vspace{-.2in}
\end{align}

\begin{table}[t]
    \footnotesize
    \centering
    \begin{tabular}{c l}
        \hline
        Notations  & {Description}  \\
        \hline
        $T_{\texttt{Load}}^{n\mathrm{L}}$, $T_{\texttt{Patch}}^{n\mathrm{L}}$  & Time to load and patch $n$ LoRA(s), respectively \\
        $T_{\texttt{Load}}^{m\mathrm{C}}$ & Time to load $m$ ControlNet(s) \\
        $T_{\texttt{Enc}}$, $T_{\texttt{VAE}}$ & Inference time of text encoder and the VAE decoder\\
        $T_{\texttt{Comp}}^{m\mathrm{C}}$ & Inference time of $m$ ControlNet(s) \\
        $T_{\texttt{Comp}}^{\mathrm{B}}$ & Inference time of the base diffusion model \\
        ${T_{\texttt{Comp}}^{\mathrm{B}^\prime}}$ & Optimized inference time of the base diffusion model \\
        $\epsilon^{n\mathrm{L}}_{\texttt{Patch}}$ &  Optimized latency for patching $n$ LoRA(s)  \\
        $\delta^{\mathrm{C}}_{\texttt{Comp}}$ & Overhead in ControlNet computation \\
        $\delta^{m\mathrm{C}}_{\texttt{Sync}}$ & Overhead in $m$ ControlNet synchronization \\
        \hline
    \end{tabular}
    \caption{Notations to model T2I inference latency.}
    \vspace{-.2in}
    \label{tab:math_symbols}
\end{table}

Our characterization identifies efficiency issues concerning ControlNet
loading ($T_{\texttt{Load}}^{m\mathrm{C}}$), sequential ControlNet execution
($T_{\texttt{Comp}}^{m\mathrm{C}}$), slow LoRA loading ($T_{\texttt
{Load}}^{n\mathrm{L}}$) and patching ($T_{\texttt{Patch}}^{n\mathrm
{L}}$), and inefficient latent batching in base model execution ($T_{\texttt
{Comp}}^{\mathrm{B}}$). We next address these issues with a holistic approach.

\section{System Design}
\label{sec:system_design}

\begin{figure}[t]
  \centering
  \includegraphics[width=0.99\linewidth]{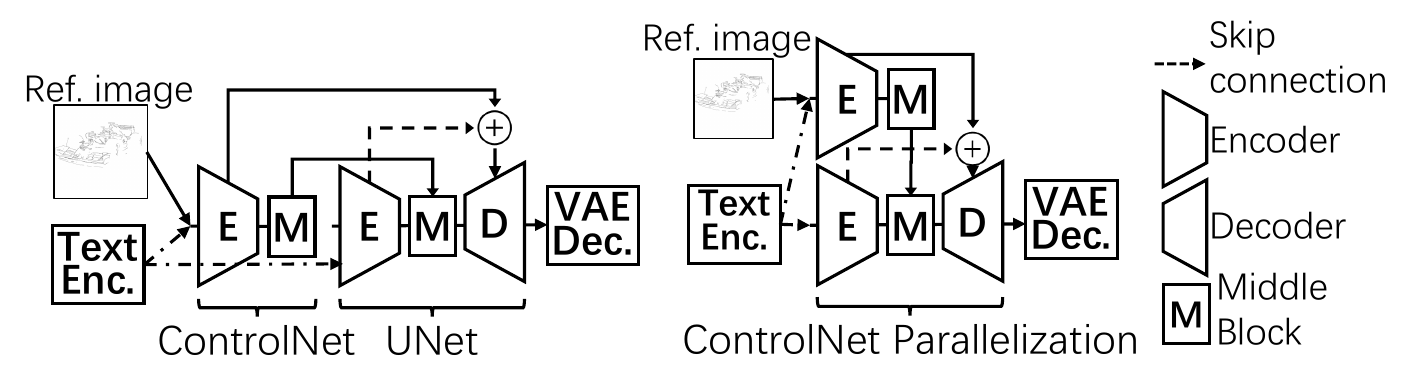}
  \caption{\textbf{Left}: Sequential ControlNet execution. \textbf{Right}: ControlNet parallelization.}
  \label{fig:controlnet_parallelization}
  \vspace{-.15in}
\end{figure}

In this section, we propose \SystemName, a system that efficiently serves the
base diffusion model and the associated add-on modules through an optimized
serving pipeline. To reduce the overhead of ControlNet loading and
computation, \SystemName introduces ControlNet-as-a-Service, which enables
ControlNet caching and parallelization in a unified design (\S\ref
{sec:design_controlnet}). To mitigate LoRA loading and patching
overhead, \SystemName overlaps LoRA fetching and base model execution at
initial denoising steps, and uses an efficient method to quickly patch LoRA
adapters (\S\ref{sec:design_lora}). \SystemName also parallelizes CFG
computation on multiple GPUs to expedite base model and ControlNet execution using
a \emph{latent parallelism} technique, together with kernel-level
optimizations (\S\ref{sec:design_unet}). Collectively, these designs
reduce the end-to-end T2I serving latency from Eq.~\eqref{eq:latency} to
\begin{align}
\vspace{-.2in}
\label{eq:our_latency}
\footnotesize
\begin{split}
T = {\epsilon^{n\mathrm{L}}_{\texttt{Patch}}} + T_{\texttt{Enc}} + \sum_{i=1}^{N} (T_{\texttt{Comp}}^{\mathrm{B} ^\prime} + \delta^{\mathrm{C}}_{\texttt{Comp}} + \delta^{m\mathrm{C}}_{\texttt{Sync}}) + T_{\texttt{VAE}}.
\end{split}
\vspace{-.3in}
\end{align}
In our evaluation, \SystemName achieves up to 7.8$\times$ reduction in latency and
1.6$\times$ throughput improvement (\S\ref{sec:eval_performance}). While we
present \SystemName primarily based on diffusion models with the UNet backbone, our
design also applies to the recently proposed diffusion transformer (DiT) models~\cite
{li2024hunyuan,sd3,peebles2023scalable}, which we elaborate in \S\ref
{sec:design_dit}.

\subsection{ControlNet-as-a-Service}
\label{sec:design_controlnet}

\PHB{Motivation.}
Our characterization study (\secref{sec:characterization_controlnet}) reveals the widespread adoption of ControlNets.
However, the existing system~\cite{diffusers} falls short in supporting ControlNets by executing them and the base diffusion model sequentially (\figref{fig:controlnet_parallelization}-Left).
With more ControlNet(s) requested, the loading and computation delay accumulates linearly (\figref{fig:lat_multi_controlnets_lora_gpu_saturation}).

\PHM{ControlNet-as-a-Service.} 
\SystemName executes ControlNet(s) on dedicated GPUs and deploys them as a separate service by decoupling ControlNet execution from the base diffusion model.
The data dependencies between ControlNet(s) and the base model allows them to execute in parallel on multiple GPUs (\figref{fig:controlnet_parallelization}-Right).
At each denoising step, the base model initiates the inference of UNet and concurrently invokes the requested ControlNet(s) by sending the reference image, latent tensor, and text embedding.
Upon completion of serving the invocation, the ControlNet(s) becomes
idle and available for the next invocation, thereby allowing a single ControlNet to be multiplexed by multiple base models.
\SystemName caches popular ControlNets in GPU memory to mitigate the loading overhead driven by the skewed distribution of ControlNet invocation (\S\ref{sec:characterization_controlnet}). ControlNet service can independently scale according to request traffic.

\PHM{ControlNet parallelization.} 
Without changing the data flow between ControlNets and the base model,  
\SystemName partitions the entire computational graph into a \emph{serial} part and a \emph{parallel} part, as shown in \figref{fig:controlnet_parallelization}.
For the UNet-base SDXL model~\cite{podell2024sdxl}, 
the \emph{serial} part consists of the one-time computation of the text encoder, one-time computation of the VAE decoder, and UNet decoder computation, while the \emph{parallel} part consists of the computations of UNet's encoder and ControlNet(s).

\SystemName distributes the computational workload in the \emph{parallel} part across multiple GPUs: the UNet is placed on one GPU while each ControlNet is allocated to a separate GPU and deployed as a service.
At each denoising step, the UNet encoder and ControlNet(s) initiate computation concurrently.
Upon completion of the UNet middle block inference, the UNet decoder \emph{synchronously} awaits the outputs from ControlNet(s) before its computation, thereby strictly preserving the original data dependencies.
Meanwhile, the ControlNet(s) becomes idle for next invocation.
This parallel computing paradigm can achieve \emph{close-to-ideal} speedup for two primary reasons.
\textbf{First}, the computational load of a ControlNet closely mirrors that of the UNet's encoder blocks and middle block, given their shared model architectural design~\cite{zhang2023controlnet}.
The only difference is that the ControlNet has additional zero convolution operators.
Consequently, their computation times are nearly uniform when executed on the same type of GPU, leading to minimal idle time and high resource utilization.
\textbf{Second}, communication between the ControlNet(s) and SDXL's UNet is minimal, \ie, 108 MiB, when using high-performance communication links, \eg, NVLink~\cite{nvlink}.
The one-time data transmission incurs a \emph{negligible} latency of less than 1 ms, ensuring efficient synchronization and minimal performance impact.

\PHM{Deploy ControlNet service.}
Given the uniformity of computational load on each GPU, the speedup gains from ControlNet parallelization on multiple GPUs primarily subject to communication overhead. 
ControlNet parallelization necessitates high-performance inter-GPU bandwidth to achieve \emph{close-to-ideal} speedup, as the outputs of ControlNets should be transmitted to the base model \emph{synchronously} to complete base model inference.

\textbf{With high inter-GPU bandwidth}, e.g., NVLink~\cite{nvlink} and InfiniBand~\cite{infiniband}, the communication overhead is negligible.
Following Gustafson’s law~\cite{Gustafson_law}, the ideal speedup of ControlNet parallelization with one ControlNet is $1.45\times$, 
while \SystemName achieves $1.42\times$ with NVLink and $1.34\times$ with InfiniBand in a Nvidia H800 cluster.
The optimality gap arises because ControlNet computation is 
$1.1\times$ longer than that of the UNet encoder and middle block.
Similar results are observed in a Nvidia A100 cluster. 

\textbf{With low inter-GPU bandwidth}, e.g., commodity Ethernet network, the communication overhead is remarkable. 
We run the SDXL model and a ControlNet, each on an AWS g5.2xlarge instances, which provides up to 10 Gbps network bandwidth.
Under these conditions, ControlNet parallelization suffers from low bandwidth, only achieving $1.1\times$ speedup.

\textbf{Summary.} The performance of ControlNet service is bounded by inter-GPU bandwidth due to the \emph{synchronous} communication between ControlNet(s) and base model. 

\PHM{ControlNet parallelization with low inter-GPU bandwidth.}
To mitigate the communication overhead, \SystemName proposes a novel scheme to \emph{pipeline ControlNet communication and base model computation}, or succinctly, \emph{pipeline ControlNet}.

\begin{figure}[t]
  \centering
  \includegraphics[width=0.49\linewidth]{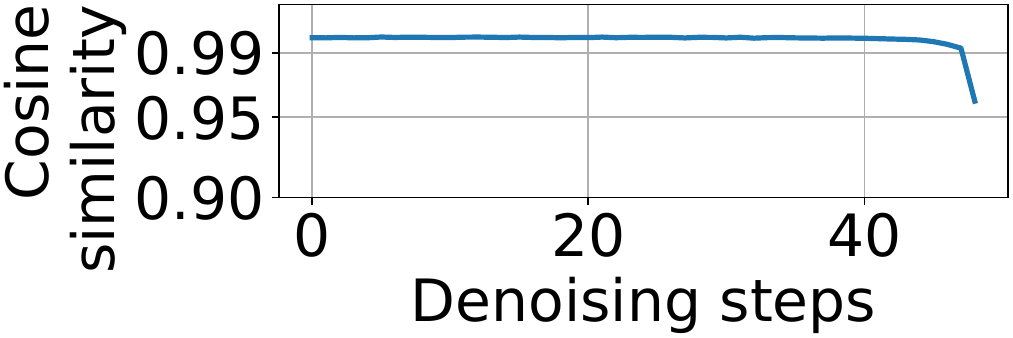}
  \includegraphics[width=0.49\linewidth]{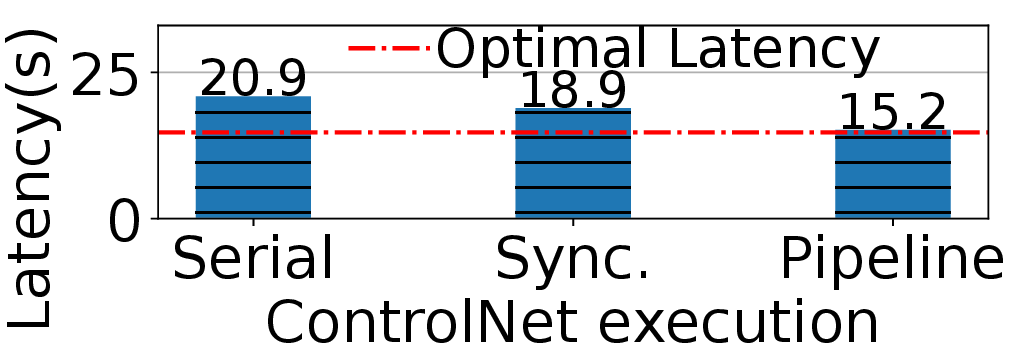}
  \caption{\textbf{Left:}  Cosine similarities of ControlNet outputs between two adjacent steps. \textbf{Right}: Latency of generating an image using various ControlNet execution with low inter-GPU bandwidth.}
  \label{fig:async_controlnet_similarity_latency}
  \vspace{-.2in}
\end{figure}

\begin{figure}[t]
  \centering
\includegraphics[width=0.99\linewidth]{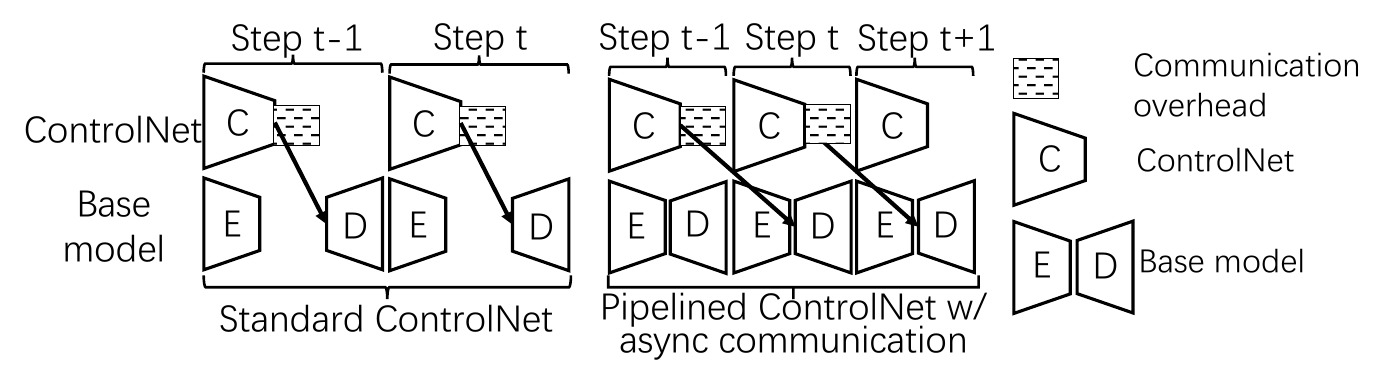}
  \caption{ControlNet parallelization with low inter-GPU bandwidth. We omit the middle blocks for simplicity.}
  \label{fig:pipeline_controlnet}
  \vspace{-.2in}
\end{figure}

We observe that the outputs of ControlNet(s) between two adjacent denoising steps are highly similar, maintaining a similarity of over 0.99 almost all the time (\figref{fig:async_controlnet_similarity_latency}-Left).
Driven by the observation, \SystemName establishes a \emph{pipeline} scheme for ControlNet parallelization with low inter-GPU bandwidth, as shown in \figref{fig:pipeline_controlnet}.
At a denoising step $t$, the base model uses the ControlNet results at step $t-1$, which have already been transmitted to the base model during the previous iteration, thereby avoiding the communication overhead.
This approach results in a $1.25\times$ speedup compared to the conventional \emph{synchronous} scheme (\figref{fig:async_controlnet_similarity_latency}-Right). 
Despite using one-step stale ControlNet results, our evaluation shows the generated images maintain comparable quality to those generated using standard workflow (\secref{sec:microbenchmark_controlnet}).

\PHM{Takeaways.} 
Within ControlNet-as-a-Service, caching ControlNets eliminates the ControlNet loading overhead ($T_{\texttt{Load}}^{m\mathrm{C}}$ in \eqnref{eq:latency}).
ControlNet parallelization reduces the $T_{\texttt{Comp}}^{m\mathrm{C}}$ in \eqnref{eq:latency} to $\delta^{\mathrm{C}}_{\texttt{Comp}} + \delta^{m\mathrm{C}}_{\texttt{Sync}}$ in \eqnref{eq:our_latency}.


\subsection{Efficient Text-to-Image with LoRAs}
\label{sec:design_lora}

\PHB{Motivation.} 
As discussed in \S\ref{sec:characterization_lora}, LoRAs are stored in a local disk or remote cache system. 
To apply a LoRA for stylizing the image generation, the system typically takes two steps.
First, it fetches the LoRA from storage and loads it into memory.
After that, it patches the LoRA to the base diffusion model by merging its weights with the parameters of the base model.
In existing system~\cite{diffusers}, the overhead of LoRA loading and patching is significant and accumulates with an increasing number of requested LoRAs (\figref{fig:lat_multi_controlnets_lora_gpu_saturation}).

\begin{figure}[t]
  \centering
\includegraphics[width=0.99\linewidth]{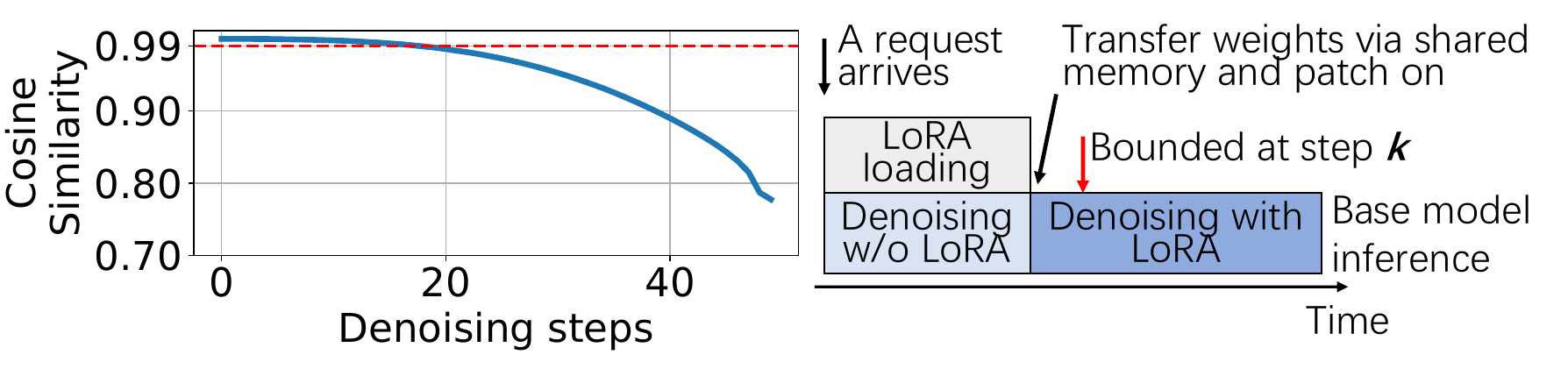}
  \caption{\textbf{Left}: Similarities between the latents generated with LoRA and those without LoRA. \textbf{Right}: Bounded asynchronous LoRA loading. }
  \label{fig:lora_feat}
  \vspace{-.2in}
\end{figure}

\begin{figure}[t]
  \centering
\includegraphics[width=0.99\linewidth]{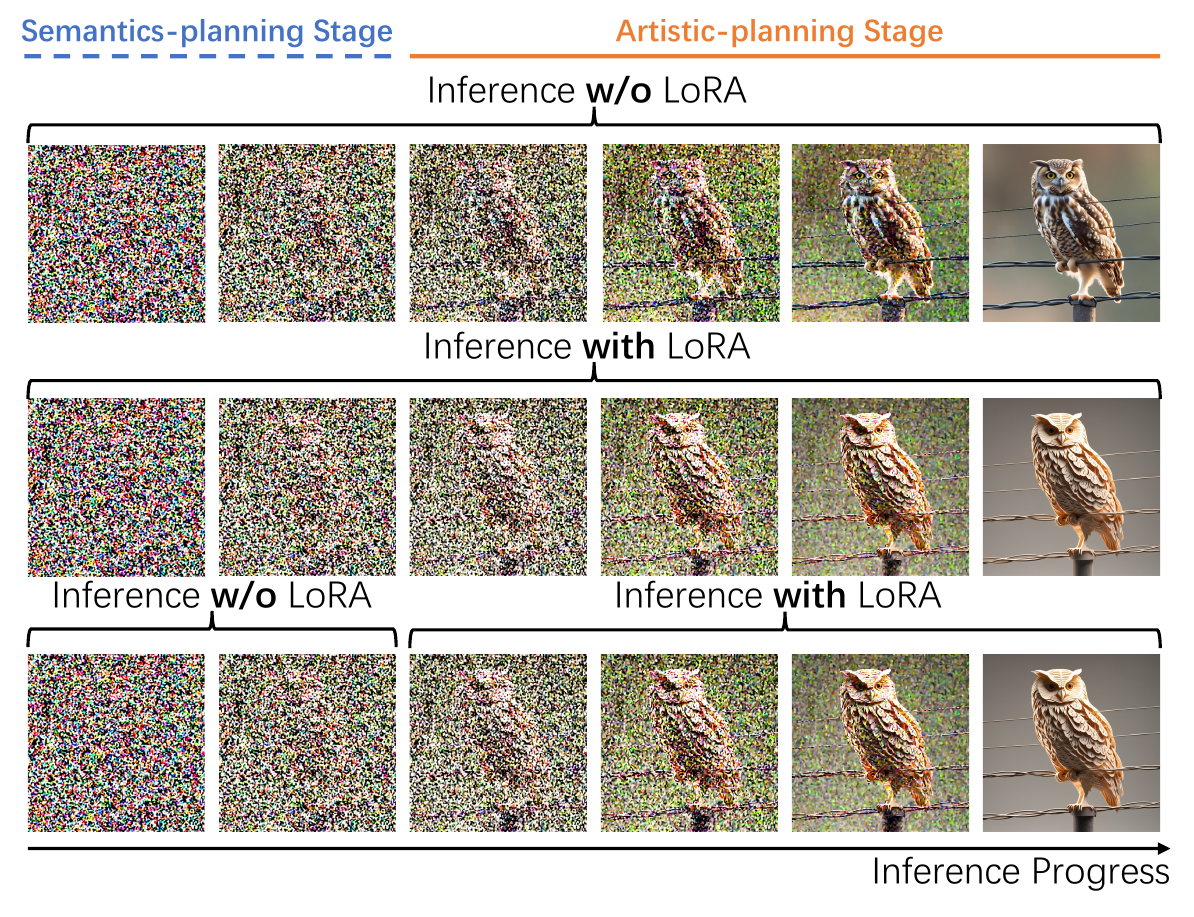}
  \caption{Images generated every 10 steps from the $0^{th}$ step using a papercut LoRA~\cite{papercut_lora}. \textbf{Prompt}: an owl standing on a wire. \textbf{Top}: W/o LoRA all the time. \textbf{Middle}: With LoRA all the time. \textbf{Bottom}: bounded async LoRA loading.}
  \label{fig:explain_lora}
  \vspace{-.2in}
\end{figure}

\PHM{Bounded asynchronous LoRA loading (BAL).}
We analyze the progress of image generation and observe that the effect of LoRA is imperceptible during the initial denoising steps.
We empirically validate this by executing the image generation process twice: once with LoRA patched on the diffusion model and once without.
We collected and calculated the cosine similarity between the latent tensors (\figref{fig:sdxl_inference}) generated with and without the LoRA at each denoising step, as demonstrated in \figref{fig:lora_feat}-Left.
The cosine similarities consistently exceed $0.99$ during the initial denoising steps, indicating that LoRA exerts \emph{minimal} effects during this stage.
After step $k$, i.e., 20, in \figref{fig:lora_feat}-Left, the similarity plunges, forming a bound that demarcates the latest possible time of LoRA patching.
\figref{fig:explain_lora} further visualizes the denoising process of image generation with and without LoRA.
The initial denoising steps constitute a semantics-planning stage~\cite{nirvana, ProSpect, si2023freeu, liu2024faster}, wherein the model determines the image composition and layout, generating visual semantics aligned with the text conditions~\cite{nirvana, ProSpect}. 
The remaining generation steps constitute an artistic-planning stage with image details gradually increasing, e.g., color, texture, and artistic style~\cite{nirvana, ProSpect}. 
Therefore, LoRA, as a method to change images' styles and details, exerts minimal effects during the semantics-planning stage, corresponding with the high similarities between the latent tensors generated with and without the LoRA in \figref{fig:lora_feat}-Left.

Driven by the observation, \SystemName proposes overlapping the LoRA loading with the initial denoising stage, as shown in \figref{fig:lora_feat}-Right.
When a request arrives, \SystemName initiates the asynchronous loading of the requested LoRA(s).
Meanwhile, it \emph{early-starts} the base stable diffusion model inference without LoRA patched.
The asynchronous LoRA loading is bounded to guarantee that image quality will not be compromised. 
Specifically, at the $k^{th}$ denoising step, if the requested LoRAs have not been loaded,
the base model inference will halt and wait until the loading completes.
Upon completion of the LoRA loading, \SystemName patches the LoRA onto the base model by merging its weights with the parameters of the base model.
The base model then proceeds with the remaining image generation process.
\SystemName executes LoRA loading in separate processes and utilizes shared memory to transfer LoRA weights from the loading processes to the base model serving process for efficient data transfer.
When serving SDXL on an H800 GPU, a LoRA with a size of 456 MiB is patched on the base model at the $11^{th}$ denoising steps on average, with a minimal overhead  (See $0C/0L$ and $0C/1L$ in \figref{fig:lat_multi_controlnets_lora_gpu_saturation}).
In cases where a request uses multiple LoRAs, \SystemName launches multiple loading processes to load the LoRAs \emph{in parallel}.

\PHM{Analysis of $k$.}
The point at which a LoRA fully loaded and patched during the denoising process influences the extent of LoRA effects manifested in the generated image. 
Delaying LoRA patching leads to a diminished presence of LoRA effects in the generated images.
\SystemName employs a profiling method to determine the optimal value of $k$.
Given a specific base model and LoRA,
we calculate the cosine
similarity between the latent tensors generated with
and without the LoRA at each denoising step, akin to those in \figref{fig:lora_feat}-Left.
Then, we assign $k$ to the step at which the similarity starts to drops below a predefined threshold, \eg, 0.99.
On H800 and A100, LoRA loading mostly completes before step $k$, effectively hiding the loading overhead.

\PHM{Efficient LoRA patching.}
Existing system~\cite{diffusers} uses the PEFT~\cite{peft} framework to merge LoRA weights with base model parameters.
For a layer in the base stable diffusion model that will be patched with LoRA, PEFT \emph{creates} a new LoRA layer to \emph{replace} the original layer in the base model.
The newly created LoRA layer augments the corresponding base model layer by incorporating LoRA weights and configurations.
However, this \emph{create\_and\_replace} operation incurs high overhead, taking 2 seconds for a LoRA of 341 MiB and occupying extra GPU memory.
Though maintaining a separate copy of LoRA weights in the new augmented layer facilitates convenient LoRA training and efficiently patching off LoRA weights after image generation, \SystemName finds it unnecessary.
First, as a serving system, \SystemName does not require the capability to support LoRA training. 
Besides, our characterization study reveals that the time interval between two consecutive requests is sufficient, \ie, longer than 1 second, to patch off LoRAs.

In \SystemName, we merge LoRA weights with base model parameters in place, obtaining two benefits.
It eliminates the latency overhead resulting from the \emph{create\_and\_replace} operation and saves GPU memory for storing separate LoRA weights.

\PHM{Takeaways.}
The designs of BAL and efficient LoRA patching reduces $T_{\texttt{Load}}^{n\mathrm{L}} +T_{\texttt{Patch}}^{n\mathrm{L}}$ in \eqnref{eq:latency} to ${\epsilon^{n\mathrm{L}}_{\texttt{Patch}}}$ in \eqnref{eq:our_latency}.

\subsection{Optimized Base Model Execution}
\label{sec:design_unet}

\PHB{Motivation.}
We now shift our focus on the base diffusion model inference time, i.e., $T_{\texttt{Comp}}^{\mathrm{B}}$ in \eqnref{eq:latency}, the last bottleneck in image generation.
In fact, when the overhead of add-on modules is minimized, the base diffusion model inference accounts for over 93\% of the end-to-end latency\footnote{For SDXL inference on an H800 GPU.}.
Moreover, the base diffusion model inference is computationally intensive, and even a small batch size of 1 can saturate high-end GPUs (\secref{sec:characterization_base_model}).
To address this challenge, \SystemName explores parallel computing to accelerate computation using multiple GPUs.
Besides, \SystemName incorporates kernel-level optimizations tailored for base diffusion model computation and its interaction with add-on modules, further enhancing performance.

\PHM{Latent parallelism for diffusion model.}
The unique CFG technique used in image generation provides an opportunity for parallel computing. 
Recalling the diffusion model computation in \secref{sec:background}, it involves denoising two latents to generate an image: one conditioned on texts and the other unconditionally. 
The two latents are then combined by
computing a weighted sum, yielding an interpolated latent.
The parallelism opportunity manifests in CFG, where the computation of denoising the two latents can be distributed across two separate GPUs, i.e., \emph{latent parallelism}.
Specifically, at each denoising step, \SystemName duplicates the latent tensor (\circled{2} in \figref{fig:sdxl_inference}) and distributes the computations of the tensors across two GPUs,
where each latent is fed into one base model on one GPU for conditioned or unconditioned denoising, respectively.
With the same GPU type, the computations are uniform, and a \emph{synchronous} communication interpolates the denoised latents through a weighted sum.

The simple yet effective latent parallelism strategy can accelerate the base diffusion model inference of a request by 1.4--1.9$\times$, depending on the GPU capabilities and base model sizes. 
The performance gain is more pronounced with larger base models and lower-end GPUs, which we will elaborate in \secref{sec:microbenchmark_unet}.
Latent parallelism incurs little overhead, because 
1) computations on different GPUs are uniform and finish at almost the same time, and
2) the communication overhead is minimal, mainly comprising the transfer of a small latent ($<$ 1 MiB).
Yet, the speedup achieved by latent parallelism may come at the expense of serving throughput when the denoising computation of a single latent does not saturate a GPU.

\PHM{Compatibility with the add-on optimizations.}
Latent parallelism can be naturally applied during ControlNet computation, since ControlNets are architecturally similar to the base models (\secref{sec:design_controlnet}). 
Therefore, both of them can benefit from latent parallelism in the ControlNet-as-a-Service design.
BAL can be seamlessly combined with latent parallelism.
If the LoRA weights are not yet loaded, we allow the base models on both GPUs to early start denoising and simultaneously patch LoRAs when loading completes.

\PHM{Kernel-level optimizations.}
\SystemName includes several kernel-level optimizations to further enhance performance, including a customized CUDA Graph implementation and specialized CUDA kernels tailored to diffusion models.
CUDA Graph is particularly suitable for T2I inference, as it use a constant batch size of 1 (\S\ref{sec:characterization_base_model}). 
Given the nearly homogeneous requested image resolutions in our production environment, we only need to maintain a small number of CUDA Graphs resident in GPU memory.
Further, we adapt the original CUDA graph to accommodate the ControlNet parallelization.
Specifically, we tailor the base model and segment it as distinct CUDA Graphs according to its data dependencies with ControlNets.
In addition to the existing optimized attention kernels~\cite{diffusers_accelerations}, \SystemName provides kernel optimizations specific to UNet-based diffusion models, including an optimized GEGLU activation operator by amalgamating GELU and matrix multiplication operations, and a fused operator that combines GroupNorm and SiLU operators to mask SiLU's overhead.

\PHM{Takeaways.}
Latent parallelism and kernel-level optimizations reduce $T_{\texttt{Comp}}^{\mathrm{B}}$ in \eqnref{eq:latency} to $T_{\texttt{Comp}}^{{\mathrm{B}^\prime}}$ in \eqnref{eq:our_latency}.

\subsection{Generalize to DiT-based Diffusion Models}
\label{sec:design_dit}

First, the design of ControlNet parallelization can be effectively generalized to DiT-based diffusion models, such as SD3~\cite{sd3} and Hunyuan-DiT~\cite{li2024hunyuan}, as the data dependencies between the ControlNet(s) and the base model in these architectures are analogous to those in UNet-based models.
This architectural similarity enables the parallel execution of ControlNet(s) and base model inference in DiT-based models, employing the similar approach as in UNet-based models.
Second, the advantages of efficient LoRA loading and patching can be effectively extended to accelerate image generation in DiT-based diffusion models, as these optimizations are agnostic to the underlying base model architecture.
Third, DiT-based models also adopt CFG in denoising and benefit from latent parallelism. 
We validate our designs with DiT-based models in \secref{sec:generalization_dit} and achieve consistent performance benefits.

\section{Implementation}
We have implemented \SystemName on top of Diffusers~\cite{diffusers}, a PyTorch-based diffusion model inference framework that integrates state-of-the-art model optimization strategies.
\SystemName is written in 5.5k lines of Python and 2.4k lines of C++/CUDA code.
ControlNet-as-a-Service, asynchronous LoRA loading, and latent parallelism are implemented in Python, while customized CUDA operators are developed in C++/CUDA.
When a request arrives, separate processes are launched to LoRA weights asynchronously and transfer the weights to the base diffusion model serving process via shared memory.


\newcommand{\NirvanaTen}{\textsc{Nirvana-10}\xspace}
\newcommand{\NirvanaTwenty}{\textsc{Nirvana-20}\xspace}
\newcommand{\Ideal}{\textsc{Ideal}\xspace}
\newcommand{\Standard}{\textsc{Diffusers}\xspace}
\newcommand{\noLora}{\textsc{NoLoRA}\xspace}

\section{Evaluation}
\label{sec:eval}

We evaluate \SystemName's performance in terms of serving latency and image quality. 
Evaluation highlights include:

\begin{itemize}[topsep=3pt, leftmargin=*, noitemsep, nolistsep, parsep=3pt, partopsep=0pt]

\item \SystemName achieves efficient serving performance without degrading image quality,  outperforming strong state-of-the-art baselines, \eg, Nirvana~\cite{nirvana} (\secref{sec:eval_performance}).

\item ControlNet-as-a-Service accelerates T2I serving with ControlNets, achieving a \emph{close-to-ideal} speedup (\secref{sec:microbenchmark_controlnet}).

\item \SystemName seamlessly incorporates LoRAs to stylize image generation while maintaining consistent serving latency (\secref{sec:microbenchmark_lora}).

\item \SystemName has a $1.7\times$ speedup in base diffusion model inference with latent parallelism and kernel optimizations (\secref{sec:microbenchmark_unet}).

\item \SystemName achieves up to $1.6\times$ throughput while using multiple GPUs compared with baselines (\secref{sec:eval_throughput}).

\item \SystemName generalizes to DiT-based diffusion models (\secref{sec:generalization_dit}).
\end{itemize}
\vspace{-.1in}

\subsection{Experimental Setup}
\label{sec:eval_setup}

\PHB{Model and serving configurations.}
We adopt SDXL~\cite{podell2024sdxl} as the primary base model for our experiments. The model and its variants have been widely used in our production cluster and benefit from comprehensive support for add-on modules.
We use the default settings to generate images, with the number of denoising steps being 50 and resolution being $1024 \times 1024$.
The ControlNets~\cite{controlnets_repo} and LoRAs~\cite{filmic_lora,papercut_lora,william_lora} used are publicly accessible in the HuggingFace repository.
We serve SDXL and ControlNets with NVIDIA H800 GPUs.

\PHM{Baselines.} 
We re-implement Nirvana~\cite{nirvana}, the SOTA text-to-image serving system, and compare it with \SystemName.
The key idea behind Nirvana is to skip the first $K$ denoising steps by utilizing a pre-cached image generated from a similar prompt to replace the randomly initialized noise latent (\figref{fig:sdxl_inference}).
By generating the image based on an intermediate representation instead of starting from scratch with noise, Nirvana aims to reduce the number of required denoising steps and improve serving latency.
In our re-implementation, we prepare the pre-cached images using the same prompts that will be used to generate the images for quality evaluation.

Including Nirvana~\cite{nirvana}, we consider the following baselines:

\begin{itemize}[topsep=5pt, leftmargin=*, noitemsep, nolistsep, parsep=3pt, partopsep=0pt] 
\item \Standard represents the standard text-to-image serving workflow incorporating ControlNets and LoRAs~\cite{diffusers}.
Images generated by \Standard adhere to the standard diffusion model inference process, which executes ControlNets sequentially and synchronously applies LoRA to the base diffusion model. 
While this approach yields images of standard quality, it incurs a relatively lengthy serving latency.

\item \NirvanaTen~\cite{nirvana} omits ten denoising steps during image generation, \ie, $K=10$.

\item \NirvanaTwenty~\cite{nirvana} aggressively skips twenty denoising steps during image generation, \ie, $K=20$.

\end{itemize}

\PHM{Metrics.}
We evaluate each baseline in terms of \emph{serving latency} and \emph{image quality}.
For serving latency, we measure the end-to-end latency of generating an image based on a given text prompt.
For image quality, we use the following quantitative metrics, which are considered essential and widely used in measuring image quality~\cite{podell2024sdxl, nirvana, ma2023deepcache,zhang2018perceptual, partiprompt}.  

\begin{itemize}[topsep=5pt, leftmargin=*, noitemsep, nolistsep, parsep=3pt, partopsep=0pt]
\item CLIP~\cite{clipscore, clip} score evaluates the alignment between 
generated images and their corresponding text prompts. A higher CLIP score indicates better alignment ($\uparrow$).

\item Fréchet Inception Distance (FID) score~\cite{fid} calculates the difference between two image sets, which correlates with human visual quality perception~\cite{nirvana}. A low FID score means that two image sets are similar ($\downarrow$).

\item Learned Perceptual Image Patch Similarity (LPIPS) score~\cite{zhang2018perceptual} quantifies the perceptual similarity between two images and has been demonstrated to closely align with human perception.
A lower LPIPS score indicates that images are perceptually more similar ($\downarrow$).

\item Structural Similarity Index Measure (SSIM) score~\cite{ssim} measures the similarity between two images, with a focus on the structural information in images. 
A higher SSIM score suggests a greater similarity between the images ($\uparrow$).

\end{itemize}

\noindent Like~\cite{nirvana, podell2024sdxl}, we conducted a user study with 75 participants to evaluate the image quality based on their visual perception. 

\PHM{Workloads.}
We use text prompts in Google's PartiPrompts (P2)~\cite{partiprompt}, which has a rich set of prompts in English.
It has both simple and complex prompts across various categories (\eg, Animals, Scenes, and World Knowledge) and challenging aspects (e.g., Detail, Style, and Imagination).
P2 has been widely used as a benchmark for image generation tasks~\cite{partiprompt, ma2023deepcache, podell2024sdxl}.
For each request, we serve it with several add-on modules, following our production trace (\tabref{tab:addon_usage}).

\begin{figure}[t]
  \centering
  \includegraphics[width=0.99\linewidth]{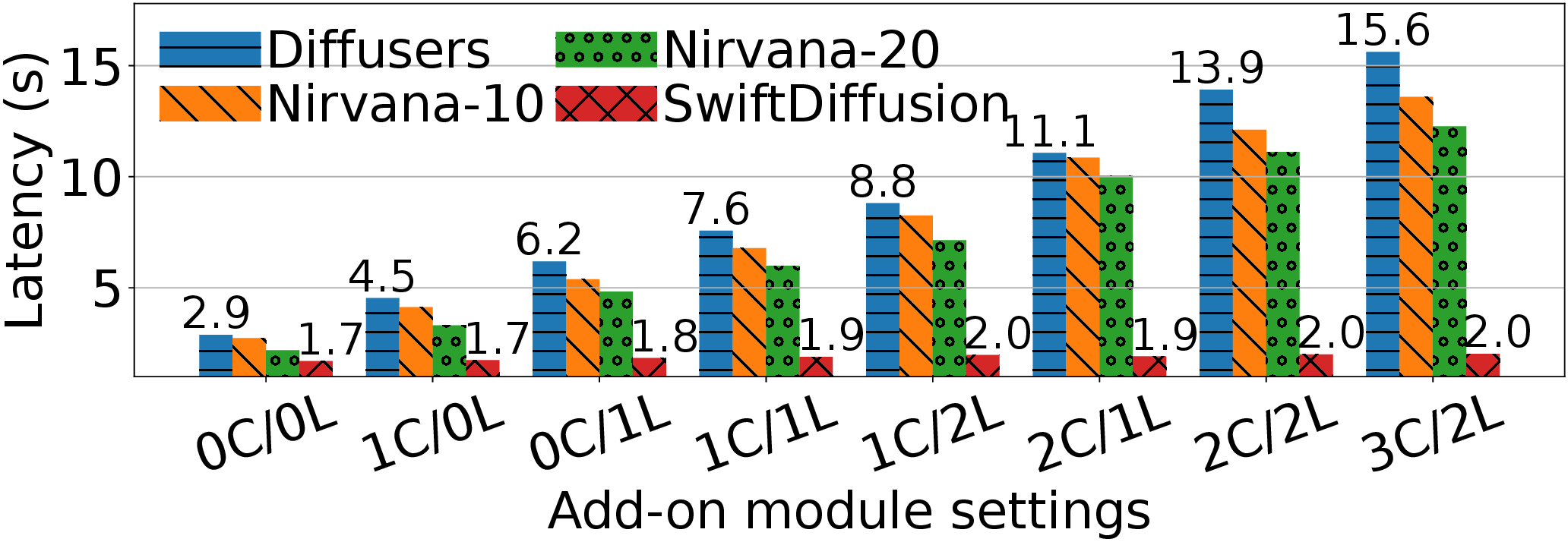}
  \caption{End-to-end serving latency with \emph{m} ControlNets and \emph{n} LoRAs (\emph{m}C/\emph{n}L).
  }
  \label{fig:end2end_latency}
  \vspace{-.2in}
\end{figure}

\subsection{End-to-End Performance}
\label{sec:eval_performance}

\PHB{Serving latency.}
We measure requests' average serving latency with different numbers of add-on modules and compare the results of each baseline in \figref{fig:end2end_latency}.
\SystemName shows its advantage across all settings that require add-on modules, achieving up to a $7.8\times$ speedup.
\SystemName outperforms other baselines by the ControlNet-as-a-Service design (\secref{sec:design_controlnet}) and efficient LoRA loading and patching (\secref{sec:design_lora}).  
In the absence of add-on modules, \SystemName achieves a 1.7$\times$ speedup compared to \Standard, due to \emph{latent parallelism} and kernel optimizations (\S\ref{sec:design_unet}).

\PHM{Image Quality.}
We compare the quality of images generated by each baseline. 
Since our ControlNets-as-a-Service design does not make any differences in the generated image content, we focus on evaluating the LoRA effects. 
Two settings are considered: the first uses a single LoRA to generate images in a papercut style~\cite{papercut_lora}, while the second employs two LoRAs to generate images in a combination of William Eggleston photography style and filmic style~\cite{william_lora, filmic_lora}.
We use the prompts in P2~\cite{partiprompt} that emphasize vivid details in images.

\begin{figure}[t]
  \centering
  \includegraphics[width=0.99\linewidth]{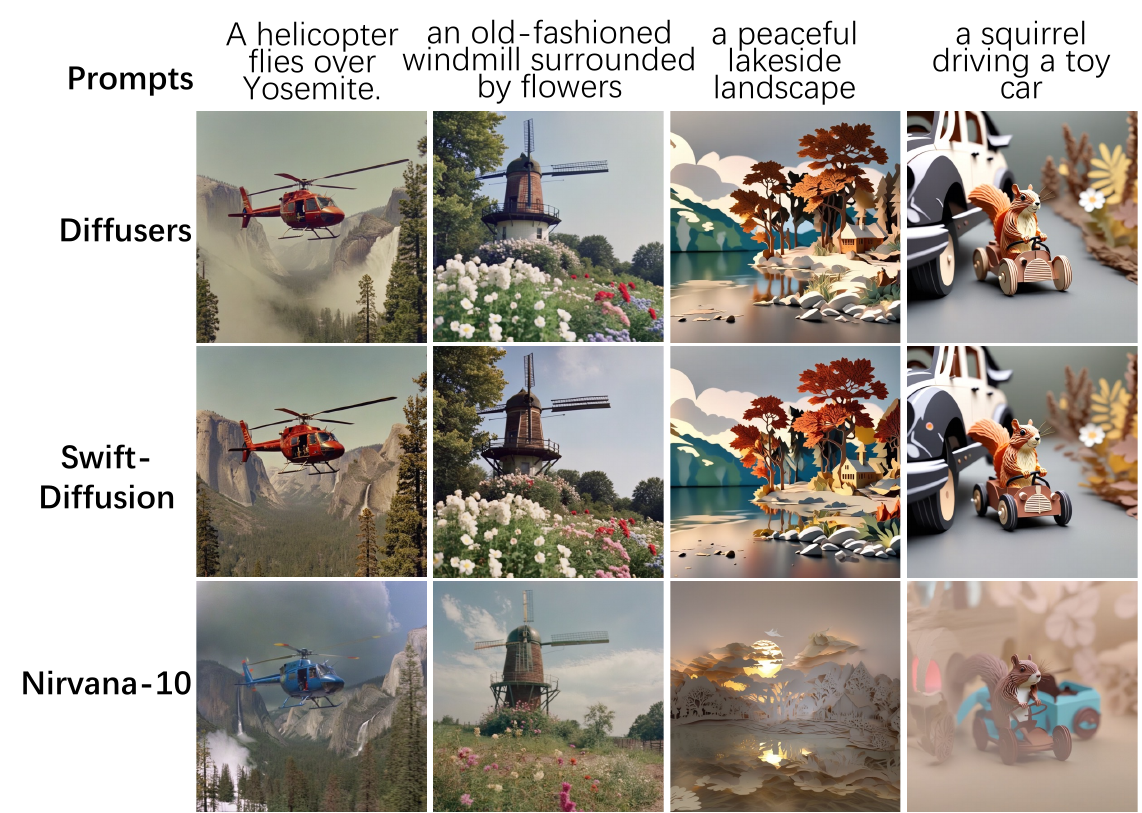}
  \caption{Real examples of generated images by \Standard~\cite{diffusers}, \SystemName, and \NirvanaTen~\cite{nirvana}.}
  \label{fig:eval_real_examples}
  \vspace{-.2in}
\end{figure}

\textit{\textbf{1)}} \textbf{Quantitative evaluation.}
\tabref{tab:image_quality_clip} shows the CLIP scores achieved by each baseline, which measure the alignment between generated images and their corresponding prompts.
The results indicate that all baselines exhibit comparable performance in terms of alignment.

\tabref{tab:image_quality} shows the FID, LPIPS, and SSIM scores achieved by each baseline. 
These metrics focus on comparing the generated images with the real images (“ground truth”).
Therefore, we use the images generated by \Standard as the ground truth, as it represents the original T2I serving workflow, while \NirvanaTen, \NirvanaTwenty, and \SystemName introduce slight modifications to accelerate the image generation.
We also consider a new baseline \noLora, which does not employ any LoRAs in image generation.
We can see that \SystemName outperforms other baselines, achieving the best performance across all metrics.
\NirvanaTen and \NirvanaTwenty fall short because they generate an image based on the contents of a cached image, which is selected only based on the prompt similarity.
Even with the same prompt, the visual contents in cached images can be drastically different (See \figref{fig:addon_effect}-Left) and may not align with the style of LoRAs.
\figref{fig:eval_real_examples} presents real examples generated by \Standard, \NirvanaTen, and \SystemName, illustrating that images generated by \Standard and \SystemName are visually almost indistinguishable, while \NirvanaTen fails to match the quality of \Standard.

\textit{\textbf{2)}} \textbf{Qualitative evaluation.}
We conducted a user study involving 75 participants to compare the quality of images generated based on human visual perception.
We consider \Standard, \NirvanaTen, and \SystemName in this part.
Inspired by Chatbot Arena~\cite{zheng2023LLMBenchmark}, we constructed an online arena that \emph{randomly} presents two images to users, offering four options: both images are acceptable, neither image is acceptable, image 1 is acceptable, or image 2 is acceptable.
Participants made their selections based on both the degree of image alignment with the prompt and their subjective aesthetic preferences.
We collected over 1.2k data points.
The findings indicate that our method is capable of producing images of the same quality as \Standard, both with a 70\% acceptance rate.
In contrast, \NirvanaTen has an overall acceptance rate below 50\% due to its skipped denoising steps and not considering the impact of add-on modules during the match process.

\begin{table}[t]
    \footnotesize
    \centering
    \begin{tabular}{c c c c c}
        \hline
        Setting    & {\Standard} & {\NirvanaTen} & {\NirvanaTwenty} & {\SystemName} \\
        \hline
        1  & 34.3  & 33.7 & 34.2 & \textbf{33.9} \\    
        \hline
        2 & 33.9  & 32.7 & 32.3 & \textbf{33.7} \\
        \hline
    \end{tabular}
    \caption{CLIP ($\uparrow$) scores.}
    \vspace{-.15in}
    \label{tab:image_quality_clip}
\end{table}

\begin{table}[t]
    \footnotesize
    \centering
    \begin{tabular}{c c c c c}
        \hline
        LoRA Setting        & Baseline  & FID ($\downarrow$) & LPIPS ($\downarrow$) & SSIM ($\uparrow$) \\
        \hline
        \multirow{3}*{ \thead{ \footnotesize{ One LoRA:} \\ \footnotesize{Papercut~\cite{papercut_lora}} } } & 
                        \noLora   & 2.71 & 0.44 & 0.59 \\
                        & \NirvanaTen  & 2.66 & 0.57 & 0.42 \\
                        & \NirvanaTwenty & 3.47  & 0.61 & 0.41 \\
                        & \SystemName  & \textbf{0.53}  & \textbf{0.26} & \textbf{0.74} \\
        \hline
        \multirow{3}*{ \thead{ \footnotesize{ Two LoRAs:} \\ \footnotesize{Filmic~\cite{filmic_lora} +} \\ \footnotesize{Photography~\cite{william_lora}} } }  & 
                        \noLora    & 1.27 & 0.45 & 0.63 \\
                        & \NirvanaTen  & 2.19 & 0.57 & 0.48 \\
                        & \NirvanaTwenty  & 2.25  & 0.62 & 0.44 \\
                        & \SystemName   & \textbf{0.78}  & \textbf{0.29} & \textbf{0.75} \\
        \hline
    \end{tabular}
    \caption{Quantitative evaluation on image quality.}
    \vspace{-.2in}
    \label{tab:image_quality}
\end{table}


\begin{figure}[t]
  \centering
  \includegraphics[width=0.49\linewidth]{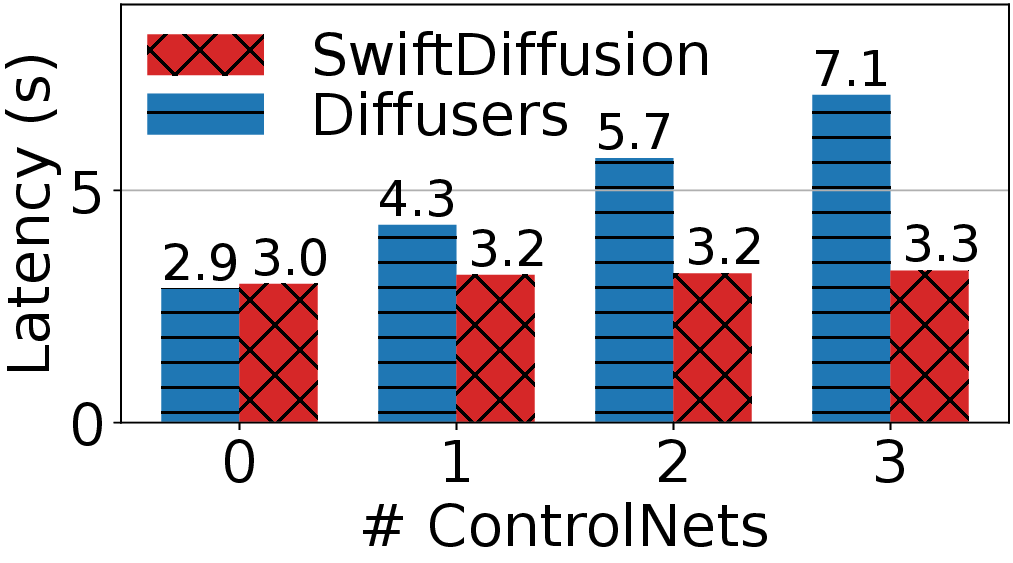}
  \includegraphics[width=0.49\linewidth]{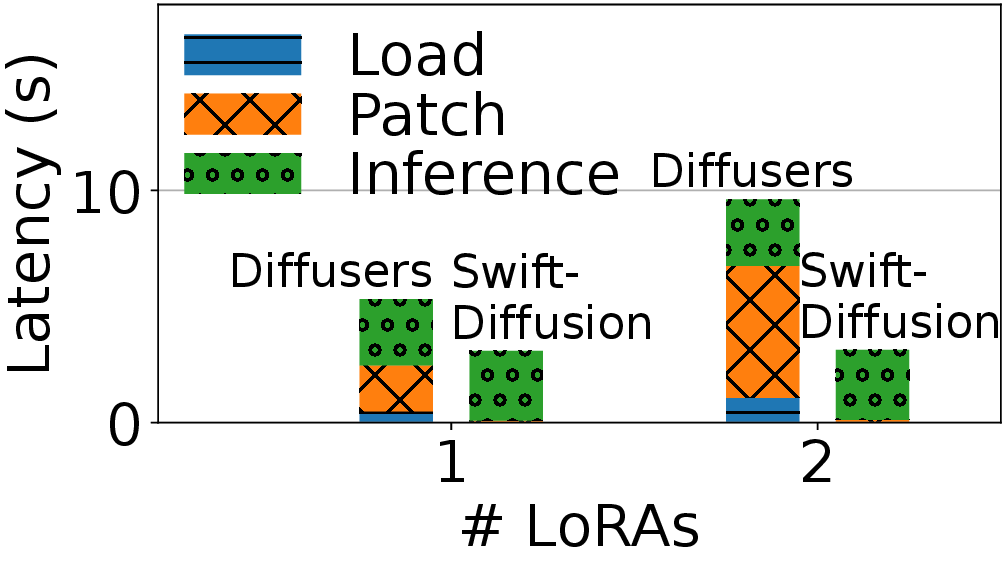}
  \caption{\textbf{Left}: Microbenchmark on ControlNets. \textbf{Right}: Microbenchmark on LoRAs.}
  \label{fig:micro_controlnet_lora}
  \vspace{-.15in}
\end{figure}

\subsection{ControlNet-as-a-Service}
\label{sec:microbenchmark_controlnet}

This section evaluates the performance of ControlNets-as-a-service design at a micro-benchmark level, isolating it from our LoRA design and optimizations in base model.
We compare \Standard and \SystemName, as Nirvana~\cite{nirvana} lacks specialized designs for ControlNets.
\figref{fig:micro_controlnet_lora}-Left illustrates the serving latency achieved by \Standard and \SystemName, where \SystemName achieves up to $2.2\times$ speedup by distributing ControlNets computation across multiple GPUs. 
Notably, \SystemName's design does not alter the image generation process, ensuring that the images generated by \Standard and \SystemName are identical.

To further analyze \SystemName's speedup, we employ Gustafson’s law~\cite{Gustafson_law}, which quantifies the theoretical speedup in execution time for a task that benefits from parallel computing.
Let $N$ denote the number of processors, and let $s$ and $p$ represent the fractions of time spent executing the serial and parallel parts of the program, \ie, $s + p = 1$.
The theoretical speedup $S$ from parallel computing is $S = s + p \times N$~\cite{wiki_gusfason}.
In the context of T2I generation with ControlNets, the serial parts comprise the computation of decoder blocks in UNet, while the parallel parts include the computation of UNet's encoder blocks together with middle block, and ControlNets (\S\ref{sec:design_controlnet}).
When using three ControlNets, the serial parts account for $s = 0.55$ and the parallel parts take $p = 0.45$, indicating a theoretical speedup of $2.35\times$. 
\SystemName's achieved $2.2\times$ speedup closely approaches this theoretical limit, demonstrating its effectiveness in leveraging parallelism across multiple GPUs.

\begin{table}[t]
    \small
    \centering
    \begin{tabular}{c c c c c}
        \hline
            {} & {CLIP ($\uparrow$)} & {FID ($\downarrow$)} & {LPIPS ($\downarrow$)} & {SSIM ($\uparrow$)} \\
        \hline
        Sync  & 34.1  & 0.66 & 0.28 & 0.75 \\    
        \hline
        Pipeline & 34.6  & 0.45 & 0.33 & 0.71 \\
        \hline
    \end{tabular}
    \caption{Quantitative evaluation on image quality.}
    \vspace{-.2in}
    \label{tab:async_image_quality}
\end{table}

\PHB{ControlNet-as-a-Service with low inter-GPU bandwidth.}
We next evaluate the latency performance and image quality of ControlNet-as-a-Service deployed over a commodity Ethernet network.
We run \SystemName on AWS g5.2xlarge instances, which provide up to 10 Gbps network bandwidth~\cite{aws_g5}.
\figref{fig:async_controlnet_similarity_latency}-Right presents the serving latency with one ControlNet, where the pipeline ControlNet achieves \emph{close-to-ideal} latency performance.
\tabref{tab:async_image_quality} shows the image quality metrics, which are derived as those in \secref{sec:eval_performance}. 
Images generated using pipeline ControlNet are comparable to those using sync ControlNet.

\subsection{Efficient T2I with LoRAs}
\label{sec:microbenchmark_lora}

This section evaluates \SystemName's design for efficient image generation with LoRAs at a micro-benchmark level, excluding other optimizations.
We also exclude Nirvana~\cite{nirvana} since it does not have specialized designs for LoRAs.
As described in \secref{sec:design_lora}, \Standard requires two steps to patch on a LoRA: first, loading the LoRA from a local disk or remote in-memory caching system, and then creating a new LoRA layer and replacing the corresponding layer in the base model to merge the LoRA weights~\cite{peft}.
This process incurs a high latency overhead, as shown in \figref{fig:micro_controlnet_lora}, with up to a $2.3\times$ increase in serving latency when using two LoRAs.
For the single LoRA case, we use a LoRA of 341 MiB.
For the two LoRA cases, we use LoRAs of 341 MiB and 456 MiB.
\figref{fig:micro_controlnet_lora}-Right shows that \SystemName's design significantly reduces the overhead of LoRA loading and patching to 230 ms through its efficient design (\S\ref{sec:design_lora}), nearly eliminating the overhead.
The image quality evaluation in \secref{sec:eval_performance} shows that \SystemName can generate high-quality images that rival those of \Standard.

\begin{figure}[t]
  \centering
  \includegraphics[width=0.49\linewidth]{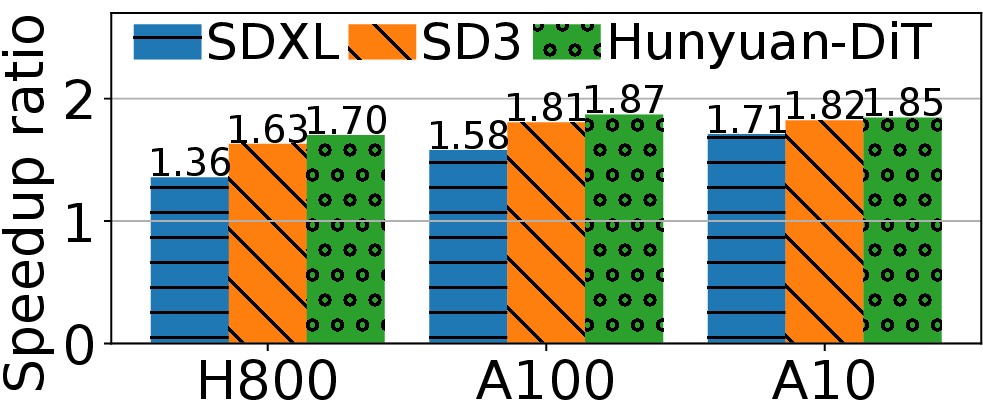}
  \includegraphics[width=0.49\linewidth]{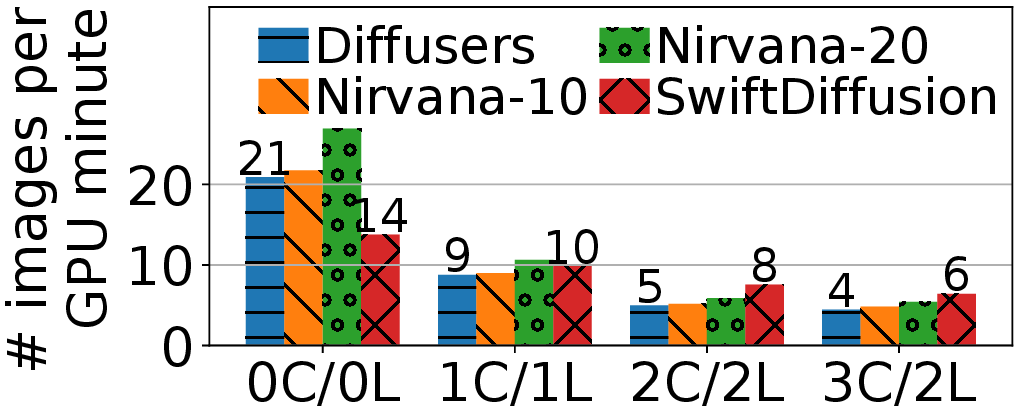}
  \caption{\textbf{Left}: Speedup ratio of latent parallelism on different GPU types and base models. \textbf{Right}: Serving throughput with \emph{m} ControlNets and \emph{n} LoRAs (\emph{m}C/\emph{n}L).}
  \label{fig:latent_parallel_speedup_throughput}
  \vspace{-.13in}
\end{figure}

\subsection{Optimize Base Model Inference}
\label{sec:microbenchmark_unet}

This section evaluates \SystemName's optimizations on the base diffusion model inference without adding any add-on modules, thereby disabling all optimizations for ControlNets and LoRAs.
As Nirvana's design only affects the number of denoising steps, we primarily compare \SystemName with \Standard.
In \secref{sec:design_unet}, we introduce two main techniques to optimize base model inference: latent parallelism and kernel-level optimization.
\figref{fig:latent_parallel_speedup_throughput}-Left shows the speedup ratio achieved by latent parallelism, where kernel-level optimizations are disabled.
The gains of latent parallelism are more pronounced on GPUs with lower compute capabilities, achieving speedup ratios of $1.36\times$, $1.58\times$, and $1.71\times$ on H800, A100, and A10, respectively.
On the basis of latent parallelism, the kernel-level optimizations, including CUDA graph and optimized kernel operators, collectively contribute to 1.24$\times$ speedup.

\subsection{Serving Throughput on GPUs}
\label{sec:eval_throughput}
\SystemName is primarily designed to minimize serving latency, but it also achieves superior serving throughput when augmented with add-on modules.
\figref{fig:latent_parallel_speedup_throughput}-Right illustrates the request serving throughput of each baseline, measured as the number of images generated per minute of GPU time.
\SystemName achieves up to a $1.6\times$ higher throughput compared to other baselines with $2C/2L$, benefiting from its efficient design of LoRA loading and patching (\S\ref{sec:design_lora}).
This demonstrates \SystemName's ability to greatly reduce serving latency (\figref{fig:end2end_latency}) while maintaining high throughput.
With $0C/0L$, \SystemName has lower throughput as a single latent computation does not saturate GPU.
\NirvanaTwenty sometimes achieves good throughput due to its aggressive design, albeit at the cost of compromising image quality.

\subsection{Generalization to DiT-based SD Model}
\label{sec:generalization_dit}
We evaluate our designs on DiT-based diffusion models.
We use SD3~\cite{sd3}, Hunyuan-DiT~\cite{li2024hunyuan}, and their add-on modules.

\PHM{ControlNet-as-a-Service} speeds up DiT-based T2I inference with ControlNet, close to the theoretical speedup gains according to the Gustafson’s law~\cite{Gustafson_law}. 
With SD3~\cite{sd3}, \SystemName achieves a $1.46\times$ speedup, rivaling the theoretically $1.50\times$ speedup.  
With Hunyuan-DiT~\cite{li2024hunyuan}, \SystemName achieves a $1.23\times$ speedup, rivaling the theoretically $1.27\times$ speedup.

\PHM{Bounded async LoRA loading.}
We evaluate the generated image quality quantitatively while enabling BAL on DiT-based models. 
We compare \SystemName and \noLora and derive the metrics as those in \secref{sec:eval_performance} in \tabref{tab:dit_image_quality}.
\SystemName consistently outperforms and generates high-quality images.

\begin{table}[t]
    \footnotesize
    \centering
    \begin{tabular}{c c c c c}
        \hline
            {} & {CLIP ($\uparrow$)} & {FID ($\downarrow$)} & {LPIPS ($\downarrow$)} & {SSIM ($\uparrow$)} \\
        \hline
        SD3  & \textbf{33.9}/33.0  & \textbf{0.07}/7.50 & \textbf{0.60}/0.65 & \textbf{0.41}/0.40 \\    
        \hline
        Hunyuan-DiT & \textbf{34.9}/34.0  & \textbf{0.12}/2.36 & \textbf{0.40}/0.49  & \textbf{0.57}/0.49  \\
        \hline
    \end{tabular}
    \caption{Image quality evaluation with \SystemName/\noLora}
    \vspace{-.2in}
    \label{tab:dit_image_quality}
\end{table}

\PHM{Latent parallelism.}
\figref{fig:latent_parallel_speedup_throughput}-Left shows the achieved speedup from latent parallelism on DiT-based models, which is consistent to that of UNet-based models.


\section{Related Works}
\label{sec:related_works}

\PHB{Model serving systems.}
Existing research on model serving systems focuses on reducing latency~\cite{miao2024SpecInfer, ye2023GRACE, yang2022INFless, crankshaw2017Clipper, wang2023Tabi}, improving throughput~\cite{ahmad2024Proteus, yang2022INFless}, enhancing performance predictability~\cite{gujarati2020Clockwork, zhang2023SHEPHERD}, and conserving resources~\cite{zhang2019MArk, wang2021Morphling, gunasekaran2022Cocktail}.
These studies concentrate on optimizing various workloads, including graph neural networks~\cite{wang2023MGG}, recommendation models~\cite{ye2023GRACE}, and large language models~\cite{wang2023Tabi, miao2024SpecInfer, yu2022Orca}.
Our work is orthogonal to the aforementioned efforts, as T2I models have drastically different computation intensity and workflow.

\PHM{T2I diffusion model inference.}
Diffusers~\cite{diffusers} is an out-of-the-box inference framework that incorporates state-of-the-art optimization strategies tailored for diffusion models.
DeepCache~\cite{ma2023deepcache} leverages the temporal consistency of high-level features to reduce redundant computations.
DistriFusion~\cite{li2024DistriFusion} facilitates the parallel execution of diffusion models across multiple GPUs.
Nirvana~\cite{nirvana} employs approximate caching to skip a certain number of denoising steps.
Yet, these works only optimize the base model inference and overlook the significant latency overhead introduced by add-on modules, which can be 4$\times$ higher than that of the base model inference.
Our work conducts a pioneering analysis of the status quo in production diffusion model deployment.
Driven by real-world traces, we address the system inefficiencies caused by incorporating add-on modules.
Notably, the existing optimizations for the base model can be seamlessly integrated.

\PHM{Serving systems with add-on modules.}
In the domain of large language models, cutting-edge research~\cite{sheng2023S-LoRA, chen2024Punica, li2024CaraServe} has proposed efficient inference techniques for models with add-on modules (\ie, LoRAs), such as CUDA-optimized operators for batched LoRA computations on GPUs and efficient GPU memory management mechanisms.
These works aim to enable multi-tenant sharing of the base model to accommodate more LoRA adapters within the same batch.
Yet, batching yields \emph{minimal} benefits in diffusion model inference, due to its compute-intensive nature.
Thus, these multi-tenant optimization strategies work ineffectively in our scenario.
In addition to LoRA, these works also overlook ControlNet, a specialized add-on module in diffusion model inference.

\section{Conclusion}
We present \SystemName, an efficient T2I serving workflow that augments image generation with ControlNets and LoRAs. 
Guided by a production characterization of commercial T2I services, \SystemName introduces three novel designs. 
Firstly, it
deploys ControlNets as a service to accelerate loading and inference. Secondly, \SystemName 
overlaps LoRA loading with base model inference and efficiently patches weights. 
Last, it optimizes the base model inference. 
Compared to existing systems, \SystemName achieves up to a $7.8\times$ latency reduction and a $1.6\times$ increased throughput  while maintaining image quality.

\bibliographystyle{plain}
\bibliography{acmart}


\end{document}